\documentclass[twocolumn]{aastex61}

\usepackage{hyperref}

\newcommand{\Msun}{M_{\odot}}
\newcommand{\Lsun}{L_{\odot}}

\newcommand{\mm}{$\mu$m}

\def\gs{\mathrel{\raise0.35ex\hbox{$\scriptstyle >$}\kern-0.6em
\lower0.40ex\hbox{{$\scriptstyle \sim$}}}}
\def\ls{\mathrel{\raise0.35ex\hbox{$\scriptstyle <$}\kern-0.6em
\lower0.40ex\hbox{{$\scriptstyle \sim$}}}}

\shorttitle{A machine-learning method for identifying SMGs}
\shortauthors{An et al.}

\begin{document}

\title{A machine-learning method for identifying multi-wavelength counterparts of submillimeter galaxies: training and testing using AS2UDS and ALESS}

\correspondingauthor{Fang~Xia An}
\email{fangxiaan@pmo.ac.cn, fangxia.an@durham.ac.uk}

\author[0000-0001-7943-0166]{Fang~Xia An}
\affiliation{Purple Mountain Observatory, China Academy of Sciences, 8 Yuanhua Road, Nanjing 210034, China}
\affiliation{University of Chinese Academy of Sciences, Beijing 100049, China}
\affiliation{Centre for Extragalactic Astronomy, Department of Physics, Durham University, Durham, DH1 3LE, UK}

\author{S.\,M.\ Stach}
\affiliation{Centre for Extragalactic Astronomy, Department of Physics, Durham University, Durham, DH1 3LE, UK}

\author{Ian Smail}
\affiliation{Centre for Extragalactic Astronomy, Department of Physics, Durham University, Durham, DH1 3LE, UK}

\author{A.\,M.\ Swinbank}
\affiliation{Centre for Extragalactic Astronomy, Department of Physics, Durham University, Durham, DH1 3LE, UK}

\author{O.\ Almaini}
\affiliation{University of Nottingham, School of Physics and Astronomy, Nottingham, NG7 2RD, UK}

\author{C. Simpson}
\affiliation{Gemini Observatory, Northern Operations Center, 670 N. A'ohuku Place, Hilo, HI 96720, USA}

\author{W.\ Hartley}
\affiliation{University of Nottingham, School of Physics and Astronomy, Nottingham, NG7 2RD, UK}

\author{D.\,T.\ Maltby}
\affiliation{University of Nottingham, School of Physics and Astronomy, Nottingham, NG7 2RD, UK}

\author{R.\,J.\ Ivison}
\affiliation{European Southern Observatory, Karl Schwarzschild Strasse 2, Garching, Germany}
\affiliation{Institute for Astronomy, University of Edinburgh, Royal Observatory, Blackford Hill, Edinburgh EH9 3HJ, UK}

\author{V.\ Arumugam}
\affiliation{European Southern Observatory, Karl Schwarzschild Strasse 2, Garching, Germany}
\affiliation{Institute for Astronomy, University of Edinburgh, Royal Observatory, Blackford Hill, Edinburgh EH9 3HJ, UK}

\author{J.\,L.\ Wardlow}
\affiliation{Centre for Extragalactic Astronomy, Department of Physics, Durham University, Durham, DH1 3LE, UK}

\author{E.\,A.\ Cooke}
\affiliation{Centre for Extragalactic Astronomy, Department of Physics, Durham University, Durham, DH1 3LE, UK}

\author{B.\ Gullberg}
\affiliation{Centre for Extragalactic Astronomy, Department of Physics, Durham University, Durham, DH1 3LE, UK}

\author{A.\,P.\ Thomson}
\affiliation{The University of Manchester, Oxford Road, Manchester, M13 9PL, UK}

\author{Chian-Chou Chen}
\affiliation{European Southern Observatory, Karl Schwarzschild Strasse 2, Garching, Germany}

\author{J.\,M.\ Simpson}
\affiliation{Academia Sinica Institute of Astronomy and Astrophysics, No.\ 1, Section 4, Roosevelt Rd., Taipei 10617, Taiwan}

\author{J.\,E.\ Geach}
\affiliation{Centre for Astrophysics Research, School of Physics, Astronomy and Mathematics, University of Hertfordshire, Hatfield AL10 9AB, UK}

\author{D.\ Scott}
\affiliation{Department of Physics and Astronomy, University of British Columbia, 6224 Agricultural Road, Vancouver, BC V6T 1Z1, Canada}

\author{J.\,S.\ Dunlop}
\affiliation{Institute for Astronomy, University of Edinburgh, Royal Observatory, Blackford Hill, Edinburgh EH9 3HJ, UK}

\author{D.\ Farrah}
\affiliation{Virginia Polytechnic Institute and State University Department of Physics, MC 0435, 910 Drillfield Drive, Blacksburg, VA 24061, USA}

\author{P.\ van der Werf}
\affiliation{Leiden Observatory, Leiden University, P.O. box 9513, NL-2300RA Leiden, the Netherlands}

\author{A.\,W.\ Blain}
\affiliation{Department of Physics and Astronomy, University of Leicester, University Road, Leicester LE1 7RH, UK}

\author{C.\ Conselice}
\affiliation{University of Nottingham, School of Physics and Astronomy, Nottingham, NG7 2RD, UK}

\author{M. J. Micha{\l}owski}
\affiliation{Astronomical Observatory Institute, Faculty of Physics, Adam Mickiewicz University, ul. Sloneczna 36, 60-286 Pozna{\' n} Poland}

\author{S.\,C.\ Chapman}
\affiliation{Department of Physics and Atmospheric Science, Dalhousie University, Halifax, NS B3H 3J5, Canada}

\author{K.\,E.\,K.\ Coppin}
\affiliation{Centre for Astrophysics Research, School of Physics, Astronomy and Mathematics, University of Hertfordshire, Hatfield AL10 9AB, UK}

\begin{abstract}
We describe the application of the supervised machine-learning algorithms to identify the likely multi-wavelength counterparts to submillimeter sources detected in panoramic, single-dish submillimeter surveys. As a training set, we employ a sample of 695 ($S_{\rm 870\mu m}\gs$\,1\,mJy) submillimeter galaxies (SMGs) with precise identifications from the ALMA follow-up of the SCUBA-2 Cosmology Legacy Survey's UKIDSS-UDS field (AS2UDS). We show that  radio emission, near-/mid-infrared colors, photometric redshift, and absolute $H$-band magnitude are effective predictors that can distinguish SMGs from submillimeter-faint field galaxies. Our combined radio\,$+$\,machine-learning method is able to successfully recover $\sim$\,85 percent of ALMA-identified SMGs which are detected in at least three bands from the ultraviolet to radio. We confirm the robustness of our method by dividing our training set into independent subsets and using these for training and testing respectively, as well as applying our method to an independent sample of $\sim$\,100 ALMA-identified SMGs from the ALMA/LABOCA ECDF-South Survey (ALESS). To further test our methodology, we stack the 870\,$\mu$m ALMA maps at the positions of those $K$-band galaxies that are classified as SMG counterparts by the machine-learning but do not have a $>$\,4.3\,$\sigma$ ALMA detection. The median peak flux density of these galaxies is $S_{\rm 870\mu m}=(0.61\pm0.03)$\,mJy, demonstrating that our method can recover faint and/or diffuse SMGs even when they are below the detection threshold of our ALMA observations. In future, we will apply this method to samples drawn from panoramic single-dish submillimeter surveys which currently lack interferometric follow-up observations, to address science questions which can only be tackled with large, statistical samples of SMGs.
\end{abstract}


\section{Introduction} \label{sec:intro}
The bulk of the population of submillimeter-luminous galaxies (SMGs) are  massive, dust-enshrouded systems which are forming stars at rates of $\gs$\,10$^{2}$--10$^{3}$\,$\Msun$\,yr$^{-1}$. At these star-formation rates (SFRs), these systems would in principle be able to form the stellar mass of massive galaxies ($M_{*}\gs$\,10$^{11}$\,$\Msun$) within just $\sim$\,100\,Myr \citep[e.g.,][]{Chapman05, Bothwell13, Casey14}. Although such strongly star-forming galaxies are  rare in the local Universe, the space density of bright SMGs (i.e., $S_{850\rm \mu m} > $\,1\,mJy, corresponding to a far-infrared luminosity, $L_{\rm IR} \gs $\,10$^{12}$\,$\Lsun$)  increases rapidly with look-back time and appears to peak at $z\sim$\,2--3 \citep[e.g.,][]{Barger99,Chapman05, Yun12, Smolcic12, Simpson14}. Due to their potentially rapid formation, SMGs have been proposed to be the progenitors of spheroidal galaxies in the local Universe \citep[e.g.,][]{Lilly99, Swinbank06, Simpson14, Simpson17}. They are also thought to be linked to quasi-stellar object (QSO) activity due the similarity of their redshift distribution to that of luminous QSOs  \citep[e.g.,][]{Coppin08}, as well as being linked to compact, red galaxies seen at $z\sim$\,1--2 \citep[e.g.,][]{Cimatti08, Whitaker12, Toft14}. These characteristics mean that SMGs may be an important stage in the formation and evolution of massive galaxies and hence are a key element to constrain models of galaxy formation and evolution. 

Submillimeter/millimeter galaxy selection benefits from the strong negative $K$-correction in these wavebands \citep{Blain93}, which enables us to detect sources above a constant flux limit and hence with near constant star-formation rates out to  high redshift ($z\sim$\,6).  In the past two decades, numerous wide-field, submillimeter surveys have been undertaken on the James Clerk Maxwell Telescope (JCMT), IRAM 30\,m, APEX, and ASTE  equipped with the SCUBA/SCUBA-2, MAMBO, LABOCA, and AZTEC cameras respectively \citep[e.g.,][and see \cite{Casey14} for a review]{Smail97, Barger98, Hughes98, Scott02, Scott12, Coppin06, Weiss09, Ikarashi11, Geach17, Wang17}. The main challenge for  follow-up studies of the sources selected from these surveys is the coarse angular resolution of the single-dish maps, with the full width at half maximum (FWHM) typically around $\sim$ 8\arcsec--10\arcsec\ at 450\,\mm\ \citep[but only for relatively small surveys,][]{Geach13, Wang17} and $\sim$\,15\arcsec--20\arcsec\ in the wide-field surveys undertake at 850--1100\,\mm\  \citep{Weiss09,Geach17} which results in uncertain identifications of the counterparts at other observed frequencies.  

Traditionally, the likely counterparts for single-dish submillimeter sources were identified by using indirect tracers of the far-infrared/submillimeter emission such as the radio, 24\,$\mu$m, or mid-infrared properties \citep[e.g.,][]{Ivison98,Smail02,Pope06,Ivison07,Barger12,Michalowski12,Cowie17}. These properties roughly track the far-infrared luminosity of galaxies and they have two additional advantages: that observations in these bands are typically at significantly higher angular resolution than the submillimeter, and that the  surface densities of sources  in these wavebands is relatively low, so that the rate of chance associations is also low. Unfortunately, the negative $K$-correction experienced in the submillimeter band  arises from the steeply rising Rayleigh-Jeans part of the spectral energy distribution (SED), the absence of which in these other wavebands means that even the deepest radio continuum or mid-infrared maps will miss the highest redshift SMGs.  Nevertheless, $\sim$\,50 percent of submillimeter sources can be located via a radio or mid-infrared identified counterpart \citep[e.g.,][]{Ivison02,Ivison07,Ivison10,Hodge13}. To improve on this situation and so construct more complete samples of SMGs it is necessary to combine a broader range of multi-wavelength properties to isolate potential SMGs from the less active galaxies which are found within the error-circles of single-dish submillimeter sources \citep[e.g.,][]{Chapin11, Alberts13, Chen16}. One additional complication of these statistical identifications is the fact that recent studies using interferometric observations in the submillimeter/millimeter  suggest that  $\gs$\,20 percent of single-dish-detected submillimeter sources actually correspond to blends of multiple SMGs \citep[e.g.,][]{Wang11, Karim13, Simpson15a, Simpson15b, Stach18a, Stach18b}. 

Recently,  interferometric observations undertaken at submillimeter/millimeter wavelengths with the Atacama Large Millimeter/submillimeter Array (ALMA) are helping to  improve our understanding of SMGs. With  angular resolution better than 1$\arcsec$, and thus sub-arcsecond positional precision, we are starting to obtain a more complete understanding of the multi-wavelength characteristics of SMGs  \citep[e.g.,][]{Hodge13, Thomson14, Swinbank14, Swinbank15, Aravena16, Walter16, Simpson17,Dunlop17,Wardlow17,Danielson17}. However, for single-dish submillimeter surveys of fields in the northern sky, it is not possible to perform ALMA follow-up, and so we must rely instead on the use of Submillimeter Array (SMA) or IRAM's Northern Extended Millimetre Array (NOEMA) to obtain interferometric identifications \citep[e.g.,][]{Hill17, Smolcic12}. Moreover, for very large samples of submillimeter sources it may be challenging to obtain complete identifications even with ALMA.

The rapid growth of data from panoramic, single-dish submillimeter surveys \citep{Geach17,Wang17, Simpson18} requires the adoption of fast, automatic techniques for identifying the likely counterparts to single-dish-detected submillimeter sources. Automatised classification using machine-learning algorithms has recently gained popularity in astronomy and has been applied to a number of problems including star/galaxy/quasars classification \citep{Bloom12, Solarz12,Malek13,Kurcz16}, or the identification of different type of supernova \citep{du15,Lochner16}.

In this work, we test two machine-learning algorithms, Support Vector Machine (SVM) and Extreme Gradient Boosting (XGBoost), to identify probable SMG counterparts from optical/near-infrared-selected galaxies. 

SVMs are a class of supervised learning algorithms based on the structural risk minimization principle developed by \cite{Vapnik95}. The main idea behind Support Vector Classification (SVC) is to determine decision planes between sets of objects with different class labels and then to calculate a decision boundary by maximising the margin between the closest points of the classes. Each single object is then classified based on its relative position in a multidimensional parameter space. 

The second machine-learning algorithm we test is XGBoost \citep{CG16}, which is a modified version of gradient boosting \citep{Friedman01} used for supervised learning problems. The basic model of XGBoost is a tree ensemble, which is a set of classification and regression trees. In this model each input feature of an object will be divided into different ``leaves'' and each ``leaf'' will be assigned a score. This score will be used as a quality on a tree structure. A greedy algorithm, that starts from a single leaf and iteratively adds branches to the tree, is used to construct structures of a tree. In this gradient boosting tree model, one of the basic functions is to search for an optimal split at each node. To make this decision, XGBoost calculates the structure score of all possible splits and find the best solution among them. In practice, multiple trees will be used together to be trained on the properties of objects in the training set and the final prediction will be made by summing the scores in the corresponding leaves of each individual tree in the tree ensemble model \citep{CG16}. 
 
Generally, there are four steps to perform a supervised machine-leaning classification: 1) construct a training set; 2) identify the optimal features that can best separate different classes; 3) train the machine-learning models to build a classifier; 4) apply to the test sample to classify the unknown objects.

In this work we exploit the multi-wavelength counterparts of $\sim$\,700 ALMA-detected SMGs identified by \cite{Stach18a,Stach18b} in their ALMA follow-up of the SCUBA-2 Cosmology Legacy Survey (S2CLS, Geach et al.\ 2017) observations of the UKIRT Infrared Deep Sky Survey \citep[UKIDSS,][]{Lawrence07} Ultra Deep Survey (UDS) field (Almaini et al.\ in prep.). We begin by identifying counterparts to ALMA SMGs by matching them to a deep $K$-band-selected photometric catalog of the UKIDSS-UDS field (Almaini et al.\ in prep; Hartley et al.\,in prep.). We then compare the multi-wavelength properties of the SMGs and a sample of non-SMG field galaxies (which lie within the footprint of our ALMA observations, but are individually undetected in these sensitive submillimeter maps) and identify those properties that can best separate these two populations. We train the machine-learning classifiers based on these selected properties to construct a method to identify probable SMG counterparts for single-dish-detected submillimeter sources that are not yet, or cannot, be observed with ALMA. By utilising our method, we can construct larger and more robust samples of counterparts to SMGs that can be used to answer the science questions related to the evolutionary cycle of SMGs and their connections with other populations.

Given the proven success of radio observations in locating counterparts to a subset of the SMG population, we adopt a two-pronged approach, where we combine a simple probability cut to select likely radio counterparts, followed by a machine-learning method applied to multi-wavelength data to increase the completeness of the resulting SMG sample. We choose to apply these two selections separately, rather than combining the radio fluxes into the machine-learning analysis, primarily because of the requirements in terms of multi-wavelength detections needed for the SVM machine-learning analysis. As we show, applying the radio and SVM machine-learning classifications independently maximises the completeness of the final SMG sample.

The plan of this paper is as follows.  We introduce the observations of the training set we use in the S2CLS UDS field and an independent test sample from the Extended {\it Chandra} Deep Field South (ECDFS) in \S\ref{s:observation}. Our methodology is described in \S\ref{s:method}. We present and discuss our results in \S\ref{s:discussion}. The main conclusions of this work are given in \S\ref{s:conclusion}. Throughout this paper we adopt a cosmology with [$\Omega_{\Lambda}$, $\Omega_M$, $h_{70}$]\,$=$\,[0.7, 0.3, 1.0]. The AB magnitude system \citep{Oke74} is used unless otherwise stated.

\section{Observational training set and test sample} \label{s:observation}

\subsection{ALMA-identified sample of submillimeter galaxies}
To construct our training set and the test sample, we employ two  wide-field, single-dish submillimeter surveys that have been uniformly followed-up using ALMA in the same submillimeter band as the original surveys (to remove ambiguity in the identification of counterparts). These then provide us with a sample of SMGs with a wide range of properties and submillimeter fluxes, and equally importantly, they yield samples of field galaxies that fall within the ALMA survey footprint but are undetected in those maps and hence can be used as a control sample of submillimeter-faint galaxies to try to distinguish the unique characteristics of SMGs. 

\subsubsection{Single-dish sample}
The UKIDSS-UDS field (RA/Dec: 02h, $-05\arcdeg$; Figure~\ref{f:multimap.eps}) was mapped with the SCUBA-2 bolometer camera \citep{Holland13} on the JCMT  at 850\,\mm\ as part of SCUBA-2 Cosmology Legacy Survey. We provide a brief overview here, the full details of observations, data reduction, and catalogue are described in \cite{Geach17}. The coverage of 0.96\,degrees$^{2}$ in UDS is relatively uniform with instrumental noise varying by only $\sim\,$5 percent across the field \citep[Fig.\,3 in][]{Geach17}. The final matched-filtered map has a noise $\le$\,1.3\,mJy\,beam$^{-1}$ rms over 0.96\,degrees$^{2}$ and of 0.82\,mJy\,beam$^{-1}$ in the deepest part. The empirical point spread function (PSF) has an FWHM of 14$\farcs$8. \cite{Geach17} identify a total of 716 submillimeter sources above a $4\,\sigma$ limit, with a false detection rate of $\sim$\,2 percent \citep[Fig.\,13 in][]{Geach17}.

We also employ a second single-dish survey sample in our analysis as an additional test of our method. This  sample comprises the 126 submillimeter sources with single-to-noise (S/N) $>3.7$ from the LABOCA ECDFS submillimeter Survey \citep[LESS;][]{Weiss09} taken with the  Atacama Pathfinder Experiment (APEX) telescope. This 870-$\mu$m map covers 0.25\,degree$^{2}$ with a 19$\farcs$2 FWHM and a 1-$\sigma$ depth of $S_{\rm 870\mu m}=$\,1.2\,mJy. The properties of this sample are thus similar to those of the S2CLS-UDS sample, but in a completely independent field with different multi-wavelength coverage and photometric selection. We refer the reader to \cite{Weiss09} for the details of these observations.

\subsubsection{ALMA follow-up} \label{s:alma_followup}

Band 7 (870\,\mm) observation have been obtained with ALMA of all the 716 submillimeter sources from the S2CLS UDS map, which are described in full in \cite{Stach18a,Stach18b}. Observations of thirty of the brightest ($S_{\rm 850\mu m} \ge $\,8\,mJy) single-dish sources were undertaken in Cycle 1 as part of a pilot project, 2012.1.00090.S \citep{Simpson15a,Simpson15b,Simpson17}, while observations of the bulk of the sample were obtained through the Cycle 3 project 2015.1.01528.S and the Cycle 4 project 2016.1.00434.S. The Cycle 1 pilot observations relied on an early interim map and in the thirty ALMA maps, 52 SMGs were detected at $\geq$\,4-$\sigma$ significance \citep{Simpson15a,Simpson15b}. However, in the final SCUBA-2 maps, three of these thirty sources fall below our sample selection criteria leaving 27 of them in our final sample of single-dish detected submillimeter sources. In Cycles 3 and 4, we observed the remaining 686 sources with $S_{\rm 850\mu m} \ge $\,3.5\,mJy from the final S2CLS map \citep{Stach18a,Stach18b}.  These observations achieve typical 1-$\sigma$ depths of $\sigma_{\rm 870\mu m}\sim $\,0.25\,mJy with synthesised beams of 0$\farcs$15--0$\farcs$3.  The ALMA maps are tapered to $\sim$\,0$\farcs$5 resolution before sources are identified.  Across all 716 single-dish submillimeter sources, we detect 695 SMGs above $>$\,4.3\,$\sigma$ (corresponding to a false detection rate of two percent).  We refer to our complete 870\,\mm\ ALMA survey of 716 SCUBA-2 sources in the UDS field as the ``AS2UDS'' survey.  We note that the ALMA primary beam of our observation is 17$\farcs$4 which encompasses the area of the SCUBA-2 beam. Full details of the observation, data reduction, source detection and cataloging are presented in \cite{Stach18a,Stach18b}.

Among the 716 ALMA maps, 108 do not contain any ALMA-identified SMG at $>$\,4.3\,$\sigma$. We label these  as ``blank-ALMA'' maps. In the remaining 608 ALMA maps, we detected 695 SMGs with fluxes from $S_{\rm 850\mu m}=$\,0.89 to 30\,mJy. In the following these maps are described as ``maps with ALMA ID''. 

The goal of this study is to develop a method to reliably and robustly identify  counterparts to single-dish-detected submillimeter sources in wide-field surveys by utilising the multi-wavelength properties of the sample of ALMA-identified SMGs. Therefore, we include the multi-wavelength galaxies lying within the 108 ``blank-ALMA'' maps in our analysis to guarantee the completeness of our parent single-dish sample. 

In our analysis we will use independent subsets of the AS2UDS SMG sample to test the reliability of our method. We also include an additional sample for this purpose:  the ALMA follow-up of the LESS survey. The ALESS survey obtained  ALMA 870-$\mu$m  observations in Cycle 0 of 122 of the 126 LESS sources \citep{Hodge13}. These early ALMA observations have a typical synthesised beam of $\sim$\,1$\farcs$6 and 1-$\sigma$ depths of $\sim$\,0.4\,mJy, but with a wider range of data quality than the later AS2UDS survey.  For this reason in this work we only use the 88 ``good quality'' ALMA maps from \cite{Hodge13} to construct our test sample.  Again these include 19 ``blank-ALMA'' maps, which lack detected SMGs. These 88 maps yield a sample of 96 ALMA-detected SMGs with multi-wavelength coverage from \cite{Simpson14}, which we will employ in our analysis. We note that the properties of this test sample differ from those of the AS2UDS sample as it is based on an IRAC-selected photometric catalog, as opposed to $K$-band for AS2UDS, and the photometric redshifts are derived using different codes in the two fields. This comparison is intended to illustrate the results which would be obtained if a training set from one field is simply applied directly to a sample selected from a second survey, with different selection and photometric coverage.

\subsection{Multi-wavelength observations}

We next describe the multi-wavelength observations of the UDS and ECDFS fields, which are used to determine the properties of our SMG samples. We will focus on the radio and redder optical and near-infrared bands, as the dusty, star-forming SMGs are expected to be typically brighter in these wavebands than the bulk of the field population \citep[e.g.,][]{Wardlow11,  Michalowski12, Hodge13, Simpson14}.

\subsubsection{VLA observations}
Since radio synchrotron emission arises from supernova remnants it provides a powerful tracer of  obscured star formation. As such radio emission has been traditionally used to identify counterparts to SMGs \citep[e.g.,][]{Ivison98, Ivison02}. 

\begin{figure}
\centering
\includegraphics[width=0.46\textwidth]{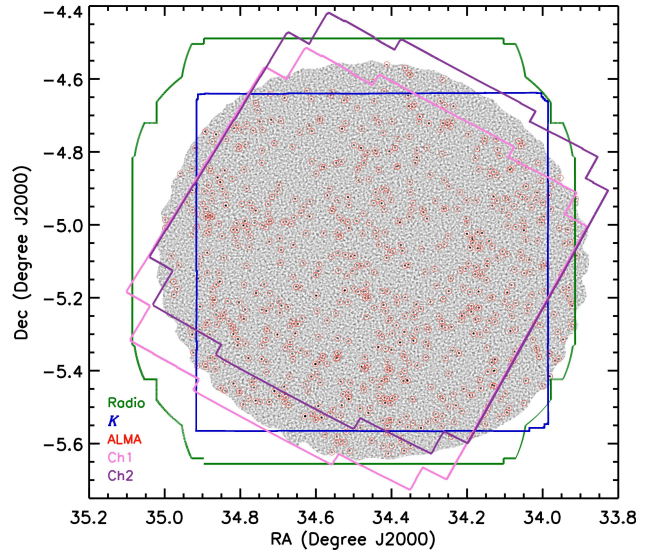}
\caption{A map showing the distribution of our ALMA survey compared to the coverage of the $K$-band, {\it Spitzer}, and VLA observations of the UDS field and overlaid on the SCUBA-2 map. We circle the positions of our 716 ALMA pointings. All but the most western two ALMA pointings are  covered by the radio map. In addition 643/716 ($\sim$\,90 percent) of the ALMA pointings fall within the deepest UKIDSS near-infrared coverage. High-quality photometric redshifts are available for those sources within the overlap region of the UKIDSS and the {\it Spitzer} IRAC 3.6\,\mm\ (Ch\,1) and 4.5\,$\mu$m (Ch\,2) imaging. There are 607/716 ($\sim$\,85 percent) of ALMA pointings in this region, which are suitable for using as a training set for our machine-learning method. We therefore limit our machine-learning analysis to this region.}
\label{f:multimap.eps}
\end{figure}

In this work, we exploit the VLA observations of the UDS at 1.4\,GHz (21\,cm), which were carried out by the UDS20 survey (Arumugam et al.\ in prep.). These VLA observations cover an area of 1.3\,deg$^{2}$ centred on the UDS field. The typical rms noise across the full VLA map is 10\,$\mu$Jy, and it is 7\,$\mu$Jy\,beam$^{-1}$ at its deepest point in the centre. In total, 6,861 radio sources are detected above $4\,\sigma$. The details of the observations, data reduction, and catalogue will be discussed in Arumugam et al.\ (in prep.). In total, 714/716 ALMA pointings fall within the VLA map (Figure~\ref{f:multimap.eps}). 
 
%
%

\subsubsection{Optical/near-infrared observations in UDS} \label{s:Multi_data}

Deep near-infrared imaging data are crucial for investigating the properties of SMGs because of their high redshifts and dusty nature. The UKIDSS-UDS represents one of the deepest near-infrared imaging surveys  over a wide area, covering 0.8\,degree$^{2}$. As shown in Figure~\ref{f:multimap.eps}, $\sim$\,90 percent (643/716) of our ALMA pointings are covered by the UKIDSS survey. 

The near-infrared image we exploit in our analysis is taken from UDS data release 11 (DR11; Almaini et al.\ in prep.), which represents the final UDS release over the whole field. Details of observations, data reduction, and catalogue extraction will be presented in the forthcoming UDS data paper (Almaini et al.\ in prep.). Briefly, the DR11 reaches 3-$\sigma$ median depths of $J=$\,26.2, $H=$\,25.7, $K=$\,25.9\,mag, which are measured in a 2$\arcsec$ diameter aperture. In total, 296,007 sources were detected from the $K$-band image  using {\sc SExtractor} \citep{Bertin96} with the photometry in the $J$ and $H$-bands  obtained in {\sc SExtractor} dual-image mode. 

The $Y$-band data are from the Visible and Infrared Survey Telescope for Astronomy (VISTA) Deep Extragalactic Observations (VIDEO) survey with 3-$\sigma$ depths of $Y=$\,25.3\,mag \citep{Jarvis13}. The optical $B$, $V$, $R_{c}$,  $i'$, and $z'$-band observations of UDS were carried out using Suprime-Cam on Subaru telescope \citep{Furusawa08} with 3-$\sigma$ depths of $B=$\,28.4, $V=$\,27.8, $R_{c}=$\,27.7, $i'=$\,27.7, and $z'=$\,26.6\,mag in 2$\arcsec$ diameter apertures. The field was also observed by the Megacam on the Canada-France-Hawaii Telescope (CFHT) in $u'$-band to a 3-$\sigma$ limiting depth of $u'=$\,27.3\,mag, again in a 2$\arcsec$ diameter aperture. 

The mid-infrared observations of the UDS were taken with IRAC and at 24\,\mm\ with MIPS by the {\it Spitzer} Legacy Program (SpUDS, PI: J.\ Dunlop). The 5-$\sigma$ depths of the  IRAC 3.6\,\mm\ and 4.5\,\mm\ observations are [3.6]\,$=$\,24.2 and [4.5]\,$=$\,24.0\,mag. 

In total, twelve-band data ($UBVRIzYJHK[3.6][4.5]$) are utilised to derive photometric redshifts for the 296,007 $K$-band-detected sources. Details of the photometric matched catalog and color measurement will be described in Hartley et al.\ (in prep.). Hartley et al.\ used {\sc EAZY} \citep[Easy and Accurate Redshifts from Yale;] []{Brammer08} to estimate the photometric redshift for the $K$-band-detected sample. To obtain unbiased and high quality photometric redshifts, they only considered those sources within the joint IRAC (SpUDS) and UKIDSS coverage and also excluded those sources that have contaminated photometry (i.e., due to halos from bright stars or other artifacts). In total, $\sim$\,85 percent (607/716) of the ALMA pointings fall in the region for which reliable photometric redshifts are available. Photometric redshifts were derived in the manner described by \cite{Simpson13} (see also \cite{Hartley13,Mortlock13}). Hartley et al.\ compare the  estimated photometric redshift of $\sim$\,6,500 sources with available spectroscopic redshifts in the DR11 and find that the accuracy of photometric redshift is $|z_{\rm spec}-z_{\rm phot}|/(1+z_{\rm spec})=$\,0.019\,$\pm$\,0.001.

\subsubsection{Multi-wavelength observations in ECDFS} \label{s:Multi_data in ecdfs}
The radio, optical and near-infrared observations of our independent test sample in ECDFS are presented in \cite{Simpson14}. The VLA 1.4\,GHz data used in \cite{Simpson14} and this work are from \cite{Miller08}. 
We use the radio catalog from \cite{Miller08} to identify radio counterparts to IRAC-based galaxies in ECDFS. \cite{Biggs11} re-reduced the VLA 1.4\,GHz imaging data in ECDFS and created a deep radio catalog containing sources down to an signal-to-noise ratio (S/N) of 3 for searching radio counterparts to single-dish-detected SMGs. We also use this deep radio catalog in our analysis to calculate the completeness of radio identification in ECDFS. The depth and quality of the multi-wavelength coverage of ECDFS is broadly  comparable to that available for UDS, in terms of number and depth of the photometric bands. For detailed information on the depth and coverage of the optical and near-infrared data in the ECDFS the reader is referred to Table~2 of \cite{Simpson14}. 

%
%
\begin{figure*}
\centering
\includegraphics[width=0.99\textwidth]{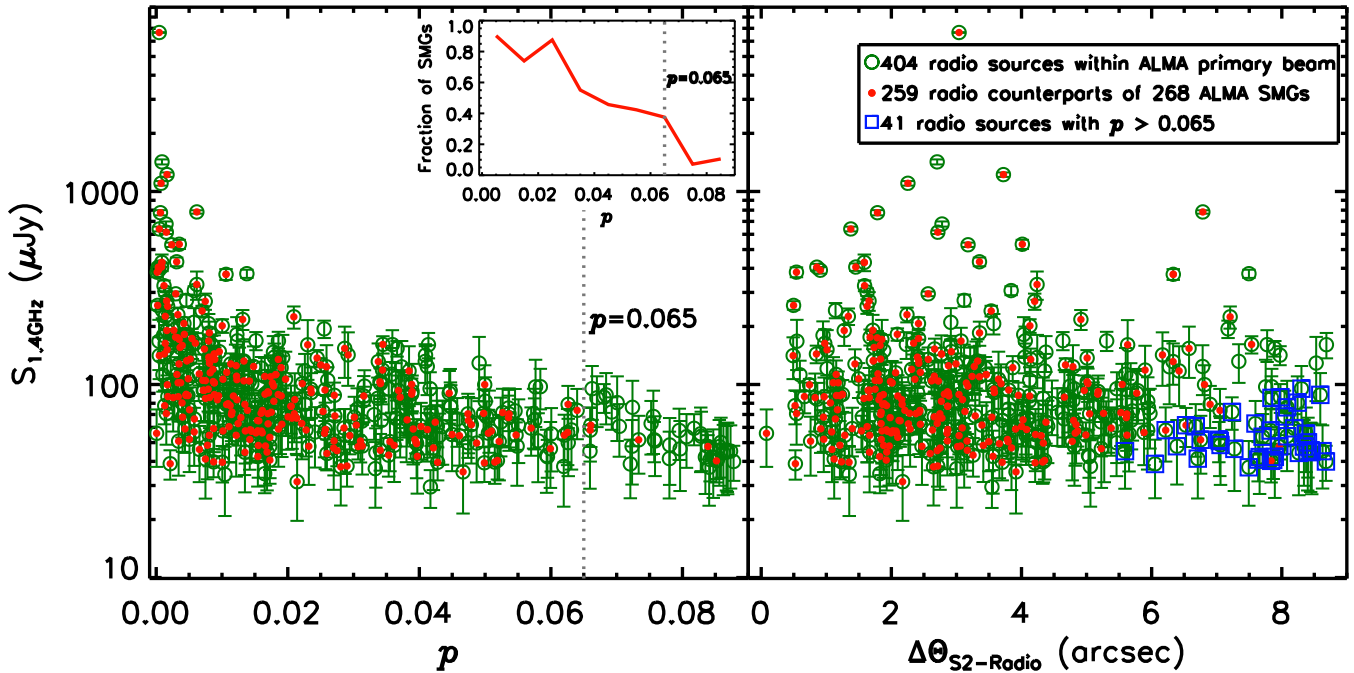}
\caption{The radio flux densities for all radio sources within the primary beams of the AS2UDS ALMA maps as a function of the corrected-Poissionian probability, $p$-value, ({\it Left}) and the offset of these radio sources from the SCUBA-2 single-dish source position ({\it Right}). In total, there are 404 radio sources within the ALMA maps in UDS (open circles). Among those, 259 radio sources are matched to 268/695 ALMA SMGs within a radius of 1$\farcs$6 (solid points), including nine ALMA SMGs which have double radio counterparts. Hence, $\sim$\,63 percent of radio sources within the ALMA maps correspond to counterparts of ALMA SMGs. We utilise the corrected-Poissonian probability, $p$-value,  to estimate the likelihood of a radio source being the counterpart of a single-dish detected submillimeter source. We show the fraction of counterparts of ALMA SMGs from all 404 radio sources within the ALMA maps as a function of $p$-value in the inset plot of the left panel. The number of counterparts of SMGs dramatically decreases when $p >$\,0.065. Therefore, we choose $p\leq$\,0.065 as a cut of ``robust'' radio  identifications in this work. There are 41 radio sources which have $p > $\,0.065 (blue squares), the majority of these are not associated with SMGs and so we adopt  $p\leq$\,0.065 as our limit for identifying radio counterparts to SMGs. Using this $p$-value, the precision of radio identification for identifying counterparts of SCUBA-2-detected SMGs is then $\sim$\,70 percent.}
\label{f:radio.eps}
\end{figure*}

\subsection{Matching SMGs to multi-wavelength data}
As the first step in our analysis we match the ALMA-identified SMGs to the multi-wavelength data from their respective fields and determine the properties of ALMA SMGs based on their multi-wavelength counterparts. 

\subsubsection{Matching to radio counterparts in UDS} \label{s:radio}
Since radio identification has been proven to be an efficient  tool to search for counterparts of bright SMGs \citep[e.g.,][]{Ivison02,Chapman05, Hodge13}, we first match our SMGs to the radio source catalogs. As shown in Figure~\ref{f:multimap.eps}, 714 of 716 ALMA maps in the UDS field are covered by the available VLA observations. There are 404 radio sources (Figure~\ref{f:radio.eps}) which fall inside the 17$\farcs$4 diameter FWHM of the primary beam coverage of the 714 ALMA maps. To identify probabilistic radio counterparts to the low-resolution, SCUBA-2-detected submillimeter sources, we include all 404 $\ge4\,\sigma$ radio sources within the ALMA  maps in our analyses.

Before matching ALMA SMGs to the radio sources, we first check the cumulative number of matches to obtain an appropriate matching radius between ALMA SMGs and radio sources. A radius of 1$\farcs$6 is chosen because the cumulative number of matches becomes flat beyond this radius. Within this matching radius, the false match rate is $\sim$\,1 percent. From the 695 AS2UDS SMGs, 693 are covered by the VLA radio observations. Among these, 268 ALMA SMGs match to 259 radio sources within 1$\farcs$6 (Figure~\ref{f:radio.eps}), with nine radio sources having two ALMA  counterparts. In total 39 percent (268/695) of AS2UDS SMGs have a radio counterparts brighter than the 4-$\sigma$ limit of the VLA catalog. 

 We then first assess the robustness of our $\ge4\,\sigma$ radio catalog. As we showed above, there are 404 radio sources in the area covered by our ALMA maps, of these 259 radio sources are counterparts to ALMA SMGs, along with 42 radio sources which lack both $K$-band and ALMA counterparts (and hence may be spurious). However, using the IRAC coverage of the field, we find that 17 of the 42 have 3.5\,\mm\ and 4.6\,\mm\ detections, indicating that about half of these are real radio sources but lack the $K$ and ALMA detections. This suggests that the spurious source fraction in our radio catalog is less than 25/404 or $\lesssim$\,6 percent. Raising the significance cut on the radio catalog to $\geq$\,5\,$\sigma$ reduces the number of $K$/IRAC/ALMA blank radio sources to 10 (from 310 radio sources, or an upper limit on the spurious fraction of $\lesssim$\,3 percent), but would also remove 40 radio-counterparts to ALMA SMGs and so reduce the completeness of our identifications. For this reason we have chosen to retain the $\geq$\,4\,$\sigma$ flux limit on the radio catalog.

To start with, for the SCUBA-2-detected submillimeter sources, we first consider all $\ge4\,\sigma$ radio sources within the ALMA primary beam as potential counterparts. Then we calculate the corrected-Poissonian probability, $p$-value \citep{Downes86, Dunlop89}, for all 404 radio sources falling in our ALMA maps by using:
\begin{eqnarray} \label{e:equation1}
E=P_{\rm c}   \qquad       P^{*} \ge P_{\rm c} \nonumber \\
E=P^{*}\{1+\ln (P_{\rm c}/P^{*})\}  \qquad  P^{*} \le P_{\rm c}
\end{eqnarray}
where $P_{\rm c}$ is the critical Poission probability level given by $P_{\rm c}=\pi r_{\rm s}^{2}N_{\rm T}$ in which $N_{\rm T}$ is the surface density of radio sources and $r_{\rm s}$ is the search radius (in this work it is the radius of ALMA primary beam). Then given $P^{*}$ for a radio source, we can derive the probability that is is a counterpart of single-dish-detected submillimeter sources by $p=\{1-\exp(-E)\}$.

As shown in Figure~\ref{f:radio.eps}, the fraction of counterparts of ALMA SMGs among the radio sources dramatically decrease when $p>$\,0.065. Hence we adopt $p\leq $\,0.065 as our limit for the probabilistic association of radio sources to single-dish submillimeter sources, while we consider those radio sources with $p =$\,0.065--0.10 as ``possible'' identifications. Looking at all 404 radio sources falling in our ALMA maps, 41 of these have $p>$\,0.065--0.10 and are thus only classed as ``possible'' counterparts (Figure~\ref{f:radio.eps}).  Of these ``possible'' counterparts, the vast majority (36/41), do not match to an ALMA-identified SMG. As a result, the five radio sources from these 41 which do match to ALMA SMGs within 1$\farcs$6 are also removed by utilising the $p$-value cut.  We also show the spatial offset of SCUBA-2 source positions and radio sources in Figure~\ref{f:radio.eps}. We see that those radio sources with $p>$\,0.065 have spatial offsets larger than 5$\farcs$5 from the nominal SCUBA-2 positions. However, if we simply adopt this smaller match radius to search for radio counterparts to SCUBA-2 sources, we will remove $\sim$\,20 of the radio counterparts to actual ALMA SMGs. Therefore, in this work, we prefer to consider all radio sources within the ALMA primary beam, but apply a $p\leq $\,0.065 cut to identify those that are likely counterparts to the SCUBA-2 detected submillimeter sources. As a result, the precision of radio identification of counterparts to single-dish-detected sources increases from 64 percent (259/404) to 70 percent (254/363) by utilising this $p$-value cut. Precision is defined as the ratio between the correctly identified SMGs and the total number of predicted SMGs by radio identification/machine-learning classification.

To identify those multi-wavelength properties that differentiate the SMGs from the wider field population, we define radio sources that do not match to an ALMA-detected SMG within 2$\farcs$6 (this is conservatively chosen to be larger than our 1$\farcs$6 matching radius) as ``non-SMG'' radio sources.  Including the 53 radio sources within the ``blank-ALMA'' maps, in total there are 137 non-SMG radio sources falling within our ALMA maps. Although, as we show later, on average the radio sources within the ``blank-ALMA'' maps have faint submillimeter emission, we put them into the sample of non-SMGs for simplicity before we perform the stacking analysis. We will discuss the properties of radio sources that are counterparts of SMGs and non-SMGs in \S\ref{s:discussion}. 

%
%
\begin{figure}
\centering
\includegraphics[width=0.49\textwidth]{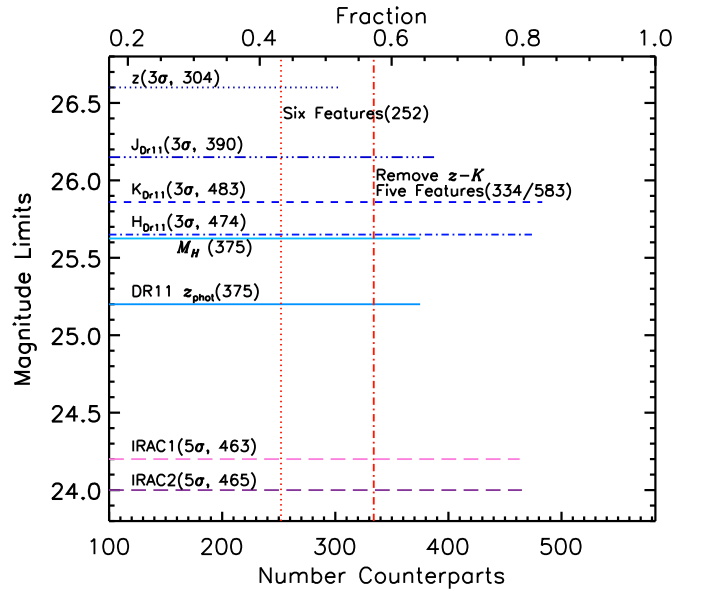}
\caption{The number of multi-wavelength counterparts to ALMA-detected SMGs within the overlap regions of UKIDSS and IRAC  coverage in the UDS field. As shown in Figure~\ref{f:multimap.eps}, $\sim$\,85 percent of our ALMA maps are covered by UKIDSS and IRAC observations, and 583/695 ALMA-detected SMGs lie in the combined footprint. The horizontal lines indicate the 3\,$\sigma$ (or 5\,$\sigma$) limit of the corresponding photometric band which is used as part of the multi-wavelength selection when identify the counterparts to SMGs. We can see that $\sim$\,83 percent of the ALMA-identified SMGs have a $K$-band counterpart, but the number of detected counterparts dramatically decreases at bluer wavelengths. We also show the number of ALMA-identified SMGs which have a photometric redshift estimate and absolute rest-frame $H$-band magnitude. The vertical lines show the fraction of SMGs which have six features (dotted line -- $(z-K)$, $(J-K)$, $(K-[3.6])$, $[3.6]-[4.5]$, $z_{\rm phot}$ and $M_H$) or five features (dot-dashed line, removing $(z-K)$) which will be used in our machine-learning method. 
}
\label{f:Number.eps}
\end{figure}

\subsubsection{Matching to near-infrared/optical counterparts in UDS} \label{s:features}
To develop a method to differentiate SMGs and non-SMGs using multi-wavelength data, we adopt the UDS DR11 photometric matched near-infrared/optical catalogue (Hartley et al.\ in prep.) to identify counterparts and measure near-infrared/optical colors of SMGs.

As we described above, only those sources within the overlapped region of UKIDSS and IRAC have sufficient photometric coverage and  estimated photometric redshifts as well as absolute magnitudes, which we will use in our machine-learning method. Hence we limit our identification of counterparts to the ALMA SMGs in this region. In total, 607/716 ALMA maps fall in this region, and 583/695 ALMA SMGs are detected within these maps with $K\leq$\,25.9\,mag. 

To select a suitable matching radius between $K$-band galaxies and ALMA SMGs, we test radii between 0$\farcs$5 and 1$\farcs$0 in steps of 0$\farcs$1 and match the $K$-band galaxies with the ALMA SMGs. At each step, we randomly offset the $K$-band galaxies in right ascension or declination by 10--20$\arcsec$ to estimate the false match fraction as a function of matching radius. At a match radius of 0$\farcs$6, 514 $K$-band galaxies from UKIDSS DR11 photometric catalog match to ALMA SMGs with a false match fraction of $\sim$\,3.5 percent ($\sim$\,18 false matches). A match radius of 0$\farcs$5 reduces the false match fraction to 2 percent ($\sim$\,10 false matches) but also reduces the total number of matches by 20. A larger match radius increases the matched sources, but the new matches are dominated by false matches. Therefore, we adopt a match radius of 0$\farcs$6.

In the overlap region of UKIDSS and IRAC, there are 483 $K$-band galaxies that match to ALMA SMGs within our adopted 0$\farcs$6 matching radius. We show the number and fraction of multi-wavelength counterparts of ALMA-detected SMGs in Figure~\ref{f:Number.eps}. We find that $\sim$\,83 percent (483/583) of the ALMA SMGs have $K$-band counterparts, but the number of counterparts dramatically decreases at bluer wavelengths due to their dusty nature (and their likely high redshifts). For the optical and near-infrared data, we use the 3-$\sigma$ limits to identify the counterparts as shown in Figure~\ref{f:Number.eps}. Because of the relatively low resolution of the IRAC data, a more conservative 5-$\sigma$ cut is adopted for identifying counterparts and measuring colors in these bands. Figure~\ref{f:Number.eps} also presents the number of SMGs that have photometric redshifts, which are estimated based on DR11 photometric catalogue, and hence have absolute $H$-band magnitudes available to be used in the following analyses. 

\subsubsection{Radio and optical/near-infrared counterparts in ECDFS}\label{s:counter_ecdf} 

The details of the identification of radio, optical/near-infrared counterparts to the ALESS SMGs in the ECDFS field are presented in \cite{Hodge13} and \cite{Simpson14} respectively. Out of the 96 ALMA SMGs, 45 have radio counterparts \citep{Hodge13}. \cite{Simpson14} measured aperture photometry in 19 wavebands for the 96 ALMA SMGs. Among these, 77 are securely detected and have sufficient photometry to derive a photometric redshift and estimate the rest-frame $H$-band absolute magnitudes. 

For the single-dish-detected submillimeter sources, we first use the IRAC-based photometric catalog of sources in ECDFS from \cite{Simpson14} to match 88 LESS submillimeter sources \citep{Weiss09} for which there are good-quality ALMA maps from \cite{Hodge13}. We include in this the 19 submillimeter sources for which the corresponding ALMA map detected no SMG (the ``blank-ALMA'' maps). In total, there are 323 IRAC-detected galaxies located within the 88 ALMA primary beams. We will use these galaxies to test our methodology in the following analysis.

\section{Method: radio + machine-learning identifications} \label{s:method}

To apply supervised machine-learning classification we require a list of observed properties for a training sample made up of submillimeter-detected and submillimeter-undetected galaxies. Therefore, firstly, we need to select those features of SMGs which best separate them from field galaxies (``non-SMGs''). Given the power of radio-identification to locate the counterparts we adopt a two-pronged approach, where we combine likelihood test to select probable radio counterparts, along with a machine-learning method to increase the completeness of the resulting SMG sample.  As we will show, we apply these two tests separately in part because of the requirements in terms of multi-wavelength detections needed for the machine-learning analysis and in part because of differences in the coverage of the field in the radio and optical and near-infrared imaging datasets.

For the machine-learning analysis, we note that previous work has shown that SMGs are in general at high redshift, are relatively bright in the rest-frame near-infrared and have red colors in optical and near-infrared wavebands \citep[e.g.,][]{Smail02,Chapman05, Hainline09, Wang12,Michalowski12, Alberts13, Simpson14, Chen16}. To compare the properties of the SMGs to the field, we use as our (``non-SMG'') control sample of those $K$-band-detected sources that are located within the  primary beams of our ALMA maps, but that are $>$\,1$\farcs$6 away from an ALMA-identified SMG. In total, there are 4,658 non-SMG $K$-band galaxies within the ALMA primary beam area (a total area of 47.3\,arcmin$^{2}$). Among them, 799 lie within the 108 ``blank-ALMA'' maps.

\subsection{``Blank-ALMA'' maps} \label{s:blank-alma}
As we described in \S\ref{s:alma_followup}, we include the 108 ``blank-ALMA'' maps in our analysis to ensure our tests
accurately reflect the success rate of identifying counterparts to ``typical'' single-dish submillimeter sources. However, due to the ambiguity about the submillimeter emission from those galaxies lying in the ``blank-ALMA'' maps, we first investigate the average far-infrared emission of these ``blank-ALMA'' maps before we include them into the sample of ``non-SMG'' galaxies used to  identify the properties that can cleanly differentiate SMGs and non-SMGs and to construct the training set for machine-learning.

We note that the false positive rate for the SCUBA-2 catalog (weighted by the number of sources at a given signal to noise) is $\sim$\,2 percent at $>4\,\sigma$ \citep{Geach17} meaning that we expect around $\sim$\,15 of our SCUBA-2 sources to be spurious, with these sources contributing to the 108 ``blank-ALMA'' maps. To test this we stack the \emph{Herschel}\,/\,SPIRE maps at the position of all 108 ``blank-ALMA'' maps. We detected significant emission with flux densities 16.4\,$\pm$\,0.6, 16.0\,$\pm$\,0.8 and 10.4\,$\pm$\,1.0\,mJy at 250, 350 and 500\,$\mu$m respectively.  Adopting the typical 850/500\,$\mu$m color for SMGs from \cite{Swinbank14} this corresponds to a typical 850\,$\mu$m flux of 3.8\,$\pm$\,0.5\,mJy, comparable to that which was detected by SCUBA-2. This indicates that the sample of ``blank-ALMA'' maps is  dominated by real submillimeter sources.

We divide the ``blank-ALMA'' maps into five bins according to their SCUBA-2 flux density to further check the influence of false positive rate of SCUBA-2 sources. We stack the SPIRE maps at the position of these maps separately and detect the emission in all SPIRE bands in all cases, with flux densities 7--20\,mJy. We also note that stacking the SPIRE images of the faintest 10 percent of the SCUBA-2 sources with ``blank-ALMA''  maps yields detections at 250 and 350\,$\mu$m. This confirms that the majority of the SCUBA-2 sources that correspond to ``blank-ALMA'' maps are real and that our estimate of 2 percent false positive sources in the parent SCUBA-2 sample is probably reasonable. The non-detection of SMGs with ALMA in these regions may due to multiplicity \citep{Hodge13, Karim13}. 

\begin{figure*}
\centering
\includegraphics[width=0.98\textwidth]{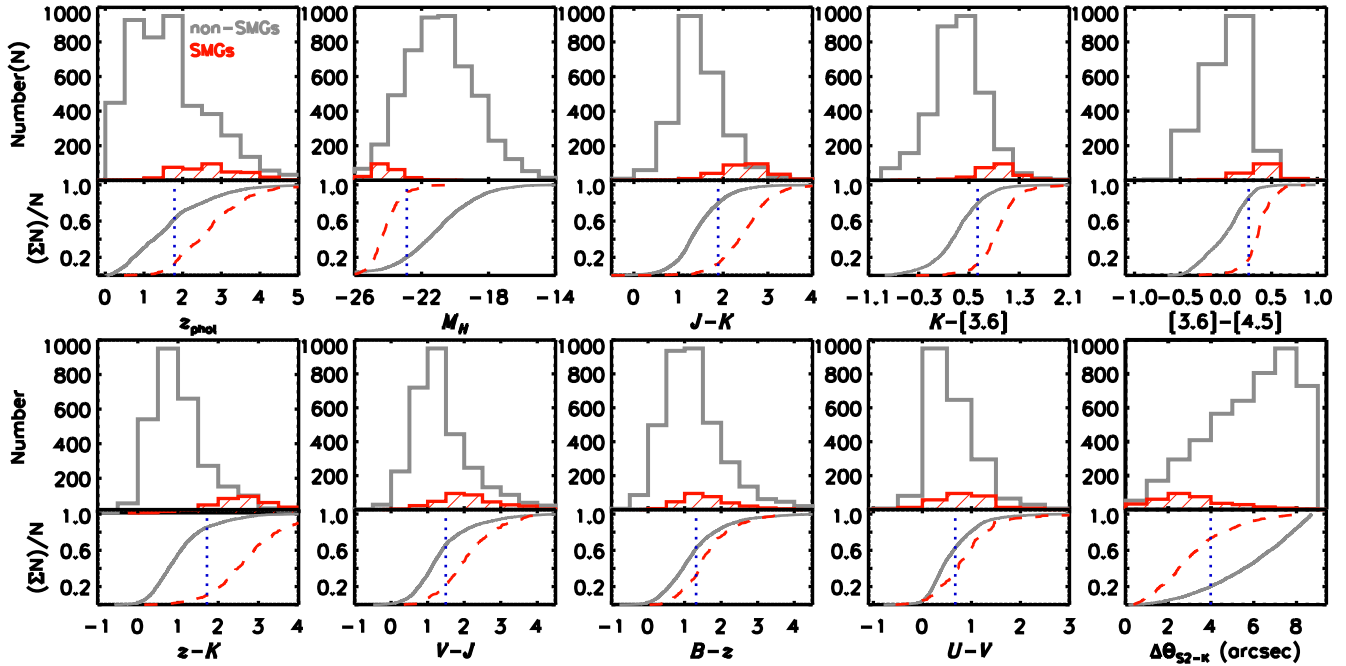}
\caption{Histograms of different observed properties of SMGs  versus non-SMG field galaxies. Non-SMGs are defined as $K$-detected galaxies that are located within the ALMA primary beams but $>$\,1$\farcs$6 away from an ALMA-detected SMGs. The distributions of all properties are normalized to the first property ($z_{phot}$) to appreciate the difference. The lower part of each panel shows the cumulative distribution and reports the Komolgorov-Smirnov (K-S) statistic for the corresponding properties. The photometric redshift, absolute $H$-band magnitude and near-infrared colors appear to have the most diagnostic power to separate these two populations,  although all of the properties have significance level of the K-S statistic  $<10^{-7}$, which means the cumulative distribution function of SMGs is significantly different from non-SMGs. The SMGs tend to lie at higher redshift, are brighter in the rest-frame $H$-band and redder in near-infrared colors. There are less distinct differences between the optical and ultraviolet color distributions for the SMGs and non-SMGs (in part because the reddest SMGs are not included in these plots). The final panel shows the spatial offset between the SCUBA-2 submillimeter sources position and $K$-band galaxies. This shows that we cannot simply use the spatial offset from the single-dish source position to classify SMG and non-SMG because of the large overlap between these two populations in terms of their spatial distributions.
}
\label{f:Color.eps}
\end{figure*}

We will show results of stacking the ``blank-ALMA'' maps at the position of machine-learning identified SMGs in \S\ref{s:discussion} which confirms that there are faint submillimeter galaxies in these maps. Therefore, to ensure a clear separation between SMG and non-SMG samples we do not include the $K$-band galaxies within the ``blank-ALMA'' maps in the ``non-SMG'' sample when identifying the characteristic properties of SMGs (Figure~\ref{f:Color.eps}) or for our training set, since they may include a disproportionate number of galaxies just below our ALMA detection limit (as we show later).

%
%

\subsection{Identifying the characteristic properties of SMGs}\label{s:identification}
Having constructed clean samples of SMGs and ``non-SMGs'', we next compare the multi-wavelength properties of these two populations to identify those properties to be used in the machine-learning analysis. We show the distributions of redshift, absolute $H$-band magnitude, optical and near-infrared colors for ALMA SMGs and non-SMG field galaxies in Figure~\ref{f:Color.eps}. We also present the results of Komolgorov-Smirnov (K-S) tests between the two populations for each of these observables. This figure demonstrates that photometric redshift, absolute $H$-band magnitude, and near-infrared colors are particularly effective at distinguishing SMGs from non-SMG galaxies. It is also clear from Figure~\ref{f:Color.eps} that those non-SMGs and SMGs detected in bluer filters show less difference in optical and ultraviolet colors -- mostly as a result of the exclusion of the redder SMGs from these plots (which require a detection in at least one of the two filters used). For this reason, previous attempts to photometrically select SMG counterparts have also focused on near-infrared color selection or optical-near-infrared (OIR) colors \citep[e.g.,][]{Smail99, Frayer04, Yun08,Michalowski12, Alberts13, Chen16}. However, as shown in Figure~\ref{f:Color.eps}, although there are clear differences between the distributions of SMGs and non-SMGs in many properties, the overlap in any individual property is substantial. Nevertheless, as we will show, the contamination from field galaxies can be efficiently reduced by combining optical/near-infrared colors,  photometric redshift and absolute rest-frame $H$-band magnitude.

The choice of which properties to use to most efficiently separate SMGs from non-SMGs for the machine-learning analysis has to balance two competing factors: precision and completeness. We have defined the precision in \S~\ref{s:radio}. 
Completeness is the number of recovered ALMA SMGs over the total number of ALMA SMGs within the overlapped region of UKIDSS and IRAC. Since including more features in the comparison is likely to yield a more precise separation, we start by using photometric redshift, absolute $H$-band magnitude ($M_H$), $(z-K)$, $(J-K)$, $(K-[3.6])$, and $([3.6]-[4.5])$ (Figure~\ref{f:Color.eps}). However, this yields a completeness of only 43 percent ALMA SMGs, which have all six of these features (as shown in Figure~\ref{f:Number.eps}). Hence to increase the completeness, we therefore remove the $(z-K)$ color which allows us to employ 57 percent of the full sample. We note that the precision of our identification is not affected by this choice since the SMGs that are red in $(z-K)$ also tend to be red in other three near-infrared colors. In fact, the precision of the identification increases by about 1 percent, which maybe be due to the enlarged sample size. 

Therefore, the features that we selected for our machine-learning classification system are:  photometric redshift ($z_{\rm phot}$), absolute $H$-band magnitude ($M_H$), and three near-infrared colors: $(J-K)$, $(K-[3.6])$, $([3.6] - [4.5])$. We find that 69 percent of the ALMA-detected SMGs lying within the UKIDSS/IRAC footprint, which have $K$-band counterparts, have secure measurement of all of  these five properties (Figure~\ref{f:Number.eps}). 

The completeness will be increased if we use fewer properties in our machine-learning analyses. However, the precision of classification decrease to just $\sim$50 percent if we only use one near-infrared color as the input feature. Therefore, we select the $K$-band detected galaxies, which have secure measurement of at least two near-infrared colors to construct the training set. The selection of photometric redshift and absolute $H$-band magnitude doesn't affect the sample size because sources with detection in three near-infrared bands (and limits/detections in the other bands) all have estimated photometric redshifts in our $K$-selected sample (Hartley et al.\ in prep.).
Removing the requirement of a secure detection at $J$-band or 4.5\mm\ modestly increases the fraction of ALMA SMGs with $K$-band counterparts which could be used for machine-learning analysis to 76 percent. In this work, we seek to develop a more complete and robust method of identifying counterparts of SMGs that are bright in several bands. This will enable us to reliably derive the physical properties of at least a subset of the SMG population. For the rest of SMGs that are only detected in the submillimeter band or that have detected counterparts in just one or two other bands, we can learn little about their physical properties.

\subsection{Radio+machine-learning identifications} \label{s:combine}

We construct a training set that includes the ALMA SMGs and non-SMG field galaxies with the selected measurements in UDS. We then train the machine-learning algorithms with these selected properties and build classifiers that can optimally distinguish the two different classes from the training set and hence predict the counterparts to the SMGs from the test sample.

\subsubsection{The machine-learning method} 
\label{s:machine-learn}
Having selected five properties that are likely to have diagnostic power to differentiate SMGs from the non-SMGs, we first use the SVM model to build a non-linear classifier for optimally separating these two populations. This is implemented by using the algorithm coded in the {\it scikit-learn}\footnote{http://scikit-learn.org} Python package \citep{Pedregosa11}. The SVC takes a labelled training set (in this case ``SMG'' versus ``non-SMG'') and associated set of feature vectors (e.g., observable colors) and attempts to build hyperplanes that maximizes the separation between the two classes in the n-dimensional ( in this case n $=5$) feature space. Having established the hyperplane(s), new, unlabelled test data can be presented to the trained classifier to determine which class it belongs to according to its relative position in this five-dimensional parameter space. 

We note that the classification can not be performed using the SVC if an object has a missing feature. This occurs if we have only a limit on the color of ($J-K$) or ([3.6] $-$ [4.5]) due to the lack of a secure detection at $J$-band or at 4.5\,\mm. Unfortunately there are a number of possible causes for the lack of $J$ or 4.5\,\mm\ detection including: dust reddening, geometry, star-formation history and redshift. Therefore, we prefer not to predict these missing values through the statistical imputation algorithms \citep[e.g.,][]{Pelckmans05}. Instead of mixing the observable properties with predicted values, we test the influence of sources with missing values using a second machine-learning model, XGBoost, which has capacity of performing classification with missing values.

We then train the SVM classifier through a training set that includes ALMA SMGs and non-SMG $K$-band galaxies, which have the secure measurement of five selected properties within the $\sim$\,50\,arcmin$^{2}$ area covered by our ALMA maps in UDS. In total, 334 ALMA SMGs and 1271 non-SMGs that have secure measurements of our five selected properties are utilised to construct the training set. 

We optimize the classifier parameters via $k$-fold cross-validation \citep{Kohavi95}. Here we use $k=5$, i.e., we randomly divide the training set into five equally sized ``folds''. The classifier is trained on $k-1$ folds and validated on the remaining fold. We  use the recovery rate (also called true positive rate, TPR, recall or sensitivity in statistics), false positive rate (FPR, also referred to as the false alarm rate or $1-$specificity), and precision (also called positive predictive value) as the evaluation metrics to optimize the parameters of the SVM classifier. The recovery rate is the ratio between the number of correctly classified SMGs and the total number of ALMA SMGs in the data set. The FPR is the number of objects  incorrectly classified as SMGs over the total number of non-SMGs in the data set. We defined the precision in \S~\ref{s:radio}, as the ratio between the number of correctly identified SMGs and the total number of predicted SMGs by the classifier. An optimized classifier will maximize the recovery rate and precision while simultaneously minimizing the FPR.

SVM classifiers use a ``kernel'' to efficiently compute the dot product between two vectors in feature space (i.e., a similarity measure) and to build a decision function which is analogous to defining a ``decision'' energy resulting from placing a kernel at the position of the observed properties of a source \citep{Cristianini00}. The five-fold cross-validation shows that the most efficient kernel function for separating SMGs from $K$-band detected galaxies is the polynomial kernel, which is defined as: 
\begin{eqnarray} \label{e:equation2}
k(x,x')=(\gamma(x \cdot x')+c_{0})^{d}
\end{eqnarray}
where $x$ and $x'$ represent feature vectors in the input space, ($x \cdot x'$) is their inner product, $d$ denotes the degree of the polynomial kernel function and $c_{0}$ is a constant coefficient which is an independent parameter in kernel function. The other two parameters of the SVM algorithms with a polynomial kernel are $\gamma$ and $C$, where $\gamma$ represents the adjustable kernel width parameter, which is responsible for the topology of the decision surface and $C$ sets the width of the margin separation different classes of objects \citep[e.g.,][]{Malek13,Kurcz16}. The five-fold cross-validation shows that the default value of these parameters in the {\it scikit-learn} package, $C=1.0$,  $\gamma=1/n$ features (here $n=5$), $d=3$ and $c_{0}=0.0$, are optimized for performing the classification in our work by SVM classifier with a polynomial kernel. 

The feature selection module in the {\it scikit-learn} package can also select the best features for classification based on the univariate statistical tests \citep{Pedregosa11}. The univariate score is derived by $U_{\rm score} = -\log (p)$, where $p$ is the $p$-value of corresponded univariate feature \citep{Pedregosa11}. Among the five features we selected, the best one for separating SMGs from non-SMGs is $(J-K)$ color with a $U_{\rm score}=$\,891, followed by $([3.6]-[4.5])$ color with a score of 707 and then $(K-[3.6])$ with a score of 695 and the absolute $H$-band magnitude has a $U_{\rm score}=$\,579. The photometric redshift has a relatively lower univariate score of 324, however, as we described above, including photometric redshift and absolute $H$-band magnitude as the input features for the machine-learning doesn't affect the completeness of our analyses but increases the recovery rate of the SVM classifier by about 6 percent.

The sample we used for performing the SVM machine-learning classification are $K$-band detected galaxies that have secure measurement of all five selected properties. To increase the completeness, we also include objects that lack a secure detection at $J$-band, i.e., have a limit on their ($J-K$) color, or lack the  detection at 4.5\,\mm, i.e., or a limit on the ([3.6] $-$ [4.5]) measurement. This increase the sample size of training set from 1605 to 1832, in which 366 are ALMA SMGs and 1466 are non-SMGs. The training set we use in our analysis is given in Table~\ref{tab:table1}.

As a test of the efficiency of the SVM classifier, we have also applied a second machine-learning classifier to our sample. This is XGBoost\footnote{https://github.com/dmlc/xgboost} \citep{CG16}, which is a scalable machine-learning system for tree boosting. In this tree ensemble model, the input features will be firstly divided into different ``leaves''. And then the algorithm computes the optimal weight of each ``leaf'' and calculates the corresponding optimal value, which will be used as a quality score of a tree structure. The structure of a tree is built by a greedy algorithm that starts from a single leaf and iteratively adds branches to the tree. Instead of enumerating all possible tree structures, XGBoost firstly calculate a gain of a ``leaf''. If the gain of corresponded leaf is smaller than the minimum loss reduction ($\gamma$), the branch will not be added to the tree. One of the key problems in tree classifiers is how to find the best split at each node (in this case ``SMG'' versus ``non-SMG''). XGBoost finds the best solution among all possible split based on the aggregated statistics according to percentiles of feature distribution. For the missing value, XGBoost classifies the instance into the optimal default direction which is learnt from the data. 
The input properties of unlabelled test data will be divided into the same leaves as the training set and the final prediction will be calculated by summing up the score in the corresponding ``leaves'' of a test object \citep{CG16}. 

Similarly, we optimize the parameters of XGBoost tree classifier via the five-fold cross-validation. Unlike the SVM implemented in the {\it scikit-learn} package, which directly predicts the class label of an object, the XGBoost classifier estimates a probability of an object being a SMG. We then also use the area-under-the-curve (AUC) of a receiver operating characteristic (ROC) curves \citep{Fawcett04} as well as the assessment metrics: recovery rate, precision and FPR, we used before to optimize the parameters of XGBoost classifier. The ROC curves are constructed by comparing the recovery rate against the FPR, as the probability threshold is varied. Typically, an AUC higher than 0.9 indicates an excellent classifier \citep[e.g.,][]{Lochner16}. 

For boosting trees, we find that a learning rate $\eta$ = 1.0 and a maximum number of iterations $num\_round=5$ are enough for performing a good classification (AUC $>0.9$). The other two parameters for a binary classification are the minimum loss reduction ($\gamma$), which is required to make a further partition on a leaf node of the tree and the maximum depth of a tree. The five-fold cross-validation indicates that $\gamma=1.0$ and the maximum depth of 6, are the optimized parameters for the XGBoost classifier. An object is classified as a SMG if the probability $\ge0.5$.

For both machine-learning algorithms, we use a uniform weight for all objects and properties. We repeat the five-fold cross-validation 100 times and calculate the median and standard deviation of each metric and present the values of these metrics for the optimized classifiers in Table~\ref{tab:table2}.

%
%
\startlongtable
\begin{deluxetable}{lrrrrr}
\tabletypesize{\scriptsize}
\tablecaption{UDS training set for machine-learning models \label{tab:table1}}
\tablehead{
\colhead{Label} & \colhead{$z_{\rm phot}$} & \colhead{$M_H$} & \colhead{($J-K$)} & \colhead{($K-[3.6]$)} & \colhead{($[3.6]-[4.5]$)}
} 
\startdata
   1\tablenotemark{a}    &    3.56  &  $-$24.59  &  2.35  &   0.73   &   0.50  \\ 
    1        &  2.50   & $-$24.05   &  2.87    &  0.96  &  0.31   \\ 
    1         &4.19    & $-$24.34    & ...        &1.25     &  0.27 \\ 
    1        & 3.10   & $-$24.22    &  3.18   &   1.16   & 0.60    \\  
    0        &  0.64    & $-$21.22  & 1.36   & $-$0.14  & $-$0.33   \\ 
    0        &  0.35   & $-$21.34    &1.46   &$-$0.48      &0.17   \\ 
    0        & 2.90    & $-$21.91   &1.49   &0.11     &0.15   \\ 
    0        &0.95    &$-$23.05    &1.88     &0.63    &$-$0.18   \\ 
    0        &0.42    &$-$18.27     & 1.16    & $-$0.68    &  ... \\
\enddata
\tablenotetext{a}{SMGs are labeled as 1 and non-SMGs are labeled as 0;}
\tablecomments{Table 1 is published in its entirety in machine-readable format.}
\end{deluxetable}

%
%
\begin{figure*}
\centering
\includegraphics[width=\textwidth]{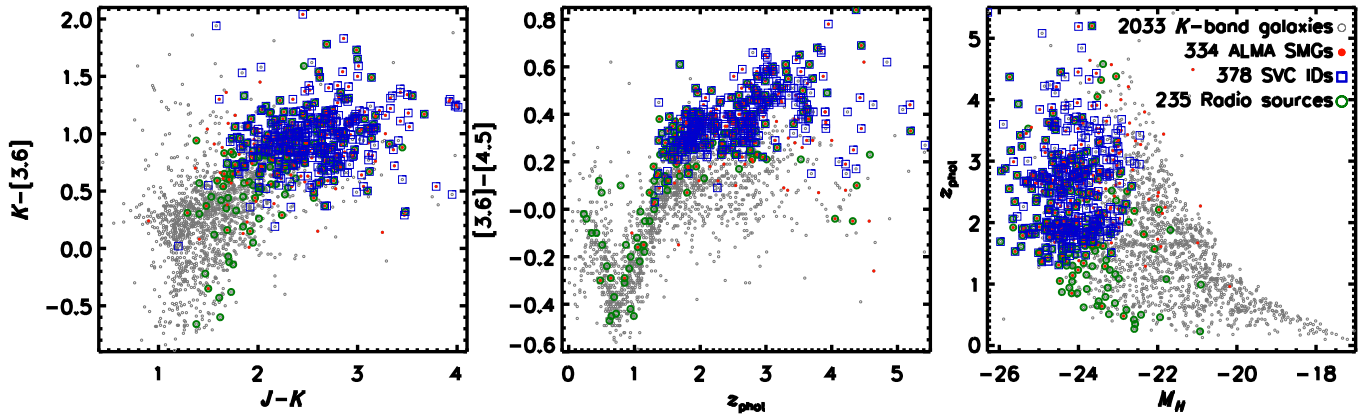}
\caption{The results of applying the support vector machine-learning classifier to identify SMGs from non-SMGs to the galaxies in the UDS field, based on a training set of the full sample of ALMA-identified SMGs in AS2UDS (termed a ``self-test''). We show the distributions of near-infrared colors, photometric redshift, and absolute $H$-band magnitude of 2,033 $K$-band-detected galaxies lying within the ALMA maps (small grey open circles). The solid points show the 334 counterparts of ALMA-detected SMGs which have secure measurement of all five observational properties. The galaxies which are classified to be counterparts of SMGs by the SVC are marked by blue open squares. We also mark those sources which have radio counterparts by large green open circles. The SVC recovers 75 percent of SMGs with a precision of $>$\,67 percent. By including radio identifications with $p \le $\,0.065, the completeness of our method reaches 85 percent with a precision of $>$\,62 percent. As we have considered all $K$-band galaxies within the ``blank-ALMA'' maps to be non-SMGs for this test, even though our stacking results show they typically have submillimeter emission just below our detection limit, the recovery rate and precision we present in the plot should be considered as lower limits.}
\label{f:self-test.eps}
\end{figure*}

\subsubsection{Test 1: self-test} \label{s:test1}
To test the efficiency of our machine-learning method, first we carry out a ``self-test'', i.e., using all $K$-band galaxies within the ALMA primary beams to build a test set. The $K$-band galaxies in the 108 ``blank-ALMA'' maps are also included in the test sample since it is not possible to know a-priori which submillimeter sources will have ``blank-ALMA'' maps (i.e., contain no SMG above a  4.3-$\sigma$ significance cut) when we identify counterparts for single-dish-detected submillimeter sources. In total, 2033 $K$-band galaxies lie within the ALMA primary beams and have secure measurements of five selected properties,  363 of these are in ``blank-ALMA'' maps. We then first utilise the training set and SVM model to identify the likely SMGs in this test sample and compare this to the actual catalog of ALMA-detected SMGs in these maps.

We present the results of the ``self-test''  in Figure~\ref{f:self-test.eps}. The SVC classifies 378 counterparts as ``SMGs''  from the 2033 $K$-band-detected galaxies within the ALMA primary beams, somewhat more than the 334 ALMA-detected SMGs in these fields. For the 334 ALMA-detected SMGs with all five features, 252/334 (75 percent) are recovered by the SVC model. The precision of this machine-learning method is therefore 67 percent (252/378). We note that this is a lower-limit on the precision for the machine-learning since we consider all $K$-band galaxies in the ``blank-ALMA'' maps as non-SMGs. However, our stacking of far-infrared observations show that there are faint SMGs present in the ``blank-ALMA'' maps and some of machine-learning method classified ``SMGs'' in the ``blank-ALMA'' maps will be true counterparts of SMGs which are marginally too faint to be detected by ALMA (as we show later). The results of the five-fold cross-validation shown in Table~\ref{tab:table2} indicate that the precision would increase to 82 percent if we had excluded the ``blank-ALMA'' maps from the analysis. 

As shown in Figure~\ref{f:self-test.eps}, those galaxies that are classified as ``SMGs'' by the SVM classifier, but that are not detected by ALMA at $>$\,4.3\,$\sigma$ (typically $S_{\rm 870\,\mu m} \ge$\,0.9\,mJy) have very similar  properties to the ALMA-detected SMGs, i.e., they are red in the near-infrared, at high-redshift, and bright in the rest-frame $H$-band. We will discuss the properties and the results of stacking the ALMA maps at the position of these galaxies in \S\ref{s:discussion}. We also note that the SMGs' counterparts that are not recovered by the machine-learning code tend to be those at lower redshifts, which are faint in the rest-frame $H$-band or blue in their near-infrared colors.

We also highlight in Figure~\ref{f:self-test.eps} those $K$-band galaxies which have radio counterparts with $p$-value $p\le$\,0.065. As we described in \S\ref{s:radio}, we use the $p$-statistic to identify radio counterparts for single-dish-detected submillimeter sources. For the 2033 $K$-band galaxies in the UDS test sample, 235 also have $>$\,4-$\sigma$ radio detections with $p\le$\,0.065.  Among these, 167/235 (71 percent) are matched to ALMA-detected SMGs within 1$\farcs$6. Therefore, half of the 334 ALMA SMGs are recovered by radio identification alone. Combining the machine-learning classification with the radio identification, 285/334 (85 percent) of the ALMA SMGs are recovered with a precision $>$\,62 percent. This proves that our combined radio and machine-learning method can  efficiently recover SMGs from the general population of $K$-band-selected galaxies.

To increase the completeness of the self-test sample, we also include the $K$-band detected galaxies that lack a secure detection at $J$-band or at 4.5\mm\ and adopt the XGBoost machine-learning module to perform the classification. The sample size is increased to 2305 with 366 of them being ALMA-detected SMGs. The XGBoost model identifies 409 ``SMGs'' from this enlarged test sample. For the ALMA SMGs, 270/366 (74 percent) are recovered with an precision of $>$ 66 percent. Combining with the radio identification, 310/366 (85 percent) of ALMA SMGs have been recovered with a precision of $>$\,62 percent. 

We note that the performances of the two machine-learning modules are very similar according to the five-fold cross-validation and this self-test (Table~\ref{tab:table2}). To keep the consistence with Figure~\ref{f:self-test.eps}, we show the analyses of the SVM classification in the following figures and use machine-learning to refer to SVM method, unless we explicitly state we are using XGBoost.

%
%
\begin{figure}
\centering
\includegraphics[width=0.48\textwidth]{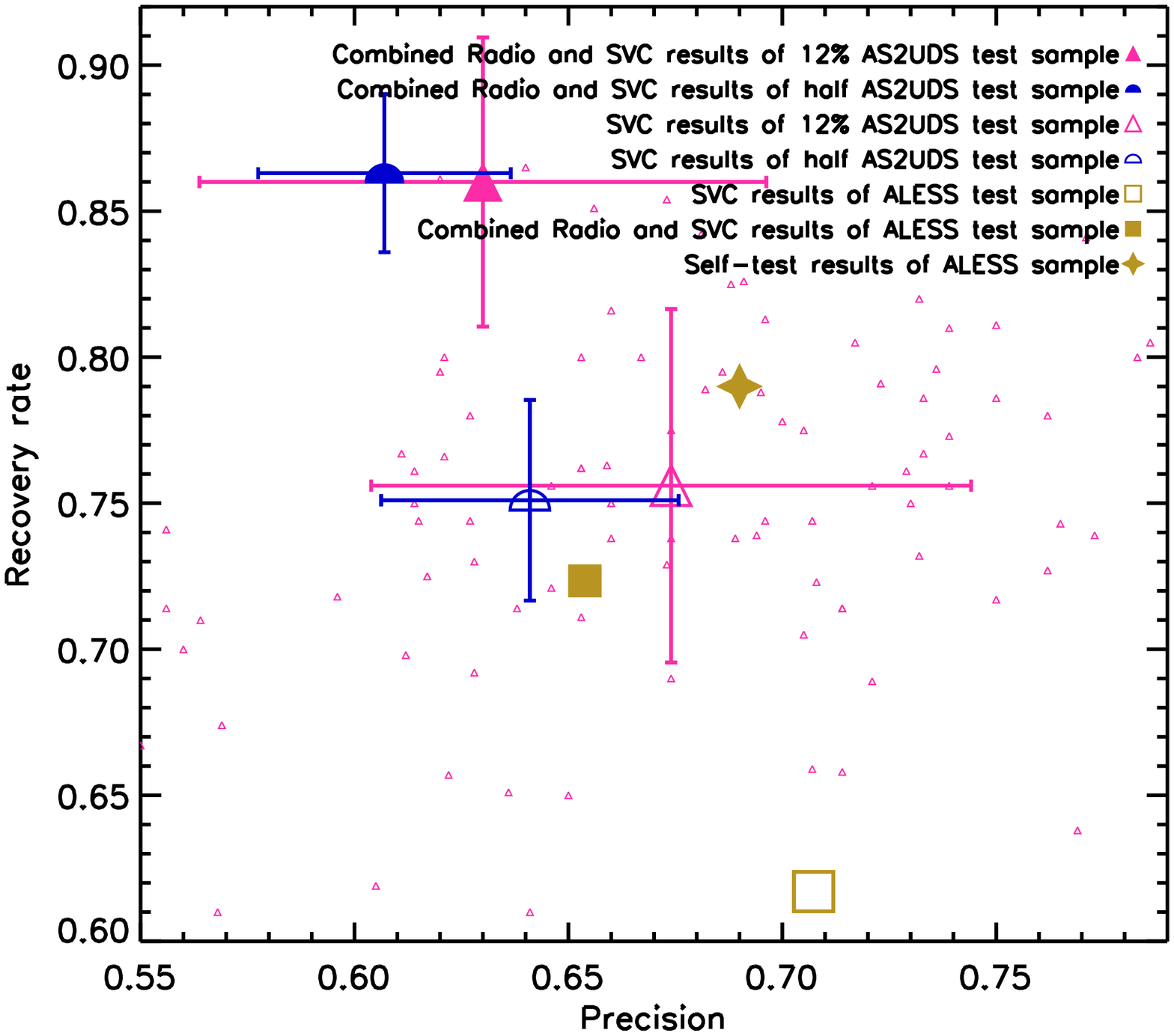}
\caption{The results of applying the support vector machine-learning method to the independent samples from the AS2UDS and ALESS surveys. The blue open half circle shows the media precision (64\,$\pm$\,3 percent) and recovery rate (75\,$\pm$\,3 percent) of our machine-learning method for the ``half--half'' test -- this involves constructing a training set from the galaxies in half of the ALMA maps in AS2UDS and testing the method on an independent sample from the other half of the ALMA maps. The filled half circle shows the results of the combined radio and machine-learning method for the ``half--half'' test.  We show error bars estimated from the variation in the derived precision and recovery rate based on 100 bootstrap simulated ``half--half'' tests. We also apply our combined radio and machine-learning method, trained on the AS2UDS sample, to the independent ALESS sample in the ECDFS field and plot this as a filled square. We recover 72 percent of ALESS SMGs with a precision of 65 percent, which can be compared to the success rate indicated by a ``self-test'' on the ALESS sample (filled star). To investigate the variation in recovery rate and precision as a function of sample size, we also randomly select sub-samples of 88 AS2UDS ALMA maps (12 percent of AS2UDS sample) matching the number of ALMA maps in ALESS. 
We show the results of recovery rate (86\,$\pm$\,5 percent) and precision (63\,$\pm$\,7 percent) of the combined radio and machine-learning method for this ``12 percent'' test sample by filled triangle, and for just the machine-learning as a large open triangle. The small open triangles represent the results of machine-learning method for 100 individual sub-tests. Four of these ``12 percent'' tests have a recovery rate as low as that seen for the ALESS test sample, while the median recovery rate of these 100 tests is same to the ``self--test'' of AS2UDS. These results illustrate the success rate of our combined radio and machine-learning method and the expected scatter in the recovery and precision when applied to smaller samples, including those selected from different fields from those used for the training set.
}  
\label{f:ecdfs1.eps}
\end{figure}

%
%
\begin{deluxetable*}{l|c|c|c||c|c|c}
\tablecaption{Summary of machine-learning/radio combined machine-learning performances for the different tests \label{tab:table2}}
\tablehead{
\colhead{Method} & \colhead{} & \colhead{Machine-learning\tablenotemark{a}} & \colhead{} & \colhead{} &  \colhead{Radio$+$Machine-learning} & \colhead{} 
} 
\startdata
  Test/metrics  &    Recovery rate & Precision  &  FPR\tablenotemark{b} &  Recovery rate  & Precision  &  FPR  \\
\hline
   SVM\tablenotemark{c}     &    ($77.2\pm4.7$)\%  & ($82.0\pm4.9$)\%  &  ($4.7\pm1.5$)\%  &  {...}  & {...}  &  {...} \\  
   XGBoost\tablenotemark{c}    &    ($76.7\pm4.9$)\%  & ($81.1\pm4.3$)\%  &  ($4.6\pm1.2$)\%  &    {...}  & {...}  &  {...} \\  
 AS2UDS self-test (SVM) &  75.4\% & 66.7\% & 7.4\% & 85.3\% & 62.2\% & 10.2\% \\
 AS2UDS self-test (XGBoost) &  73.8\% & 66.0\% & 7.2\% & 84.7\% & 62.2\% & 9.7\% \\
  AS2UDS ``half-half'' test &  ($75.1\pm3.4$)\% & ($64.1\pm3.5$)\% & ($8.3\pm1.2$)\% & ($86.3\pm2.7$)\% & ($60.7\pm3.0$)\% & ($10.1\pm1.9$)\%   \\
  12 percent AS2UDS test &  ($75.6\pm6.1$)\% & ($67.4\pm7.0$)\% & ($8.3\pm1.2$)\% & ($86.0\pm5.0$)\% & ($63.0\pm6.6$)\% & ($10.1\pm1.9$)\%   \\
  ALESS test &  61.7\% & 70.7\% & 6.5\% & 72.3\% & 65.4\% & 9.7\% \\
  ALESS self-test &  72.3\% & 73.9\% & 6.5\% & 78.7\% & 68.5\% & 9.2\% \\
\enddata
\tablenotetext{a}{The machine-learning refers to the SVM unless we state that we used XGBoost;}
\tablenotetext{b}{False positive rate (FPR) which is defined as the number of objects that are incorrectly classified as SMGs over the total number of non-SMGs in the data set;}
\tablenotetext{c}{The results of five-fold cross-validation for the optimized machine-learning classifier.}
\end{deluxetable*}

\subsubsection{Test 2: independent test} \label{s:test2}

We expect the ``self--test'' will provide an overly optimistic indication of the success rate of our method as it uses the same sample for both the training and testing.  For that reason we also undertake a number of independent tests, which use distinct samples for the training and testing.  

Firstly, we divide the AS2UDS sample into independent halves to test our method, which we will term a ``half--half'' test.  We randomly assign the $K$-band galaxies in half the ALMA maps to the training set and use the galaxies within the other half of the ALMA maps as the test sample. We then utilise this training set and our combined radio and SVC machine-learning method to classify the likely counterparts of SMG in the independent test sample.  We repeat this ``half--half'' test 100 times to estimate the scatter in the recovery rate and precision. The median recovery rate of the combined radio and machine-learning method of these tests is 86\,$\pm$\,3 percent with a media precision of 61\,$\pm$\,3 percent (Figure \ref{f:ecdfs1.eps}). This ``half--half'' test confirms the success of our method when used to identify the counterparts of SMGs using a training set with similar photometric coverage and depth. This success rate is therefore expected to be representative of that which will be achieved when we apply our method to identify the SMG counterparts in the S2COSMOS survey \citep{Simpson18}.

The next independent test we perform is to apply the trained SVM classifier to the independent sample of ALMA-identified SMGs in the ECDFS field from the ALESS survey \citep[][]{Hodge13,Simpson14}. As we described in \S\ref{s:counter_ecdf}, there are 323 IRAC-selected galaxies located within the 88 ALMA primary beams.  232/323 of these galaxies have secure measurement of the five selected properties which are used in our SVM classification. Among these, 47/232 sources match to ALESS MAIN sample SMGs within 1$\farcs$5.

We show the results from applying the SVM classifier, trained on the AS2UDS sample, to the identification of the SMGs in ALESS in Figure~\ref{f:ecdfs1.eps}. The recovery rate of the machine-learning is 62 percent with a precision 
$>$\,71 percent. As we have included the 19 ``blank-ALMA'' maps in our test sample (which include some galaxies classified as non-SMG, but which are actually just below our submillimeter flux limit), we believe this precision is a lower limit. We also match these 232 IRAC-based galaxies with radio catalog from \cite{Miller08}. Again, the radio identification can recover half of ALMA SMGs with a precision of 75 percent. Hence, combining the radio identification and machine-learning classification, we recover 72 percent of ALMA SMGs with a lower-limit on the precision of 65 percent. 

We note that the recovery rate for ALESS sample is lower than that achieved in either the ``self-test'' or ``half--half'' tests on the AS2UDS sample. To understand the cause of this we also carry out a ``self-test'' on the ALESS sample (i.e., we use the ALESS SMGs and non-SMGs as both the training and analysis samples) and find that the recovery rate of classification increases from 72 to 79 percent while the precision increases to 69 percent. The recovery rate of ALESS SMGs is still lower than that of ``self-test'' on the AS2UDS SMGs. 

\begin{figure}[h!]
\centering
\includegraphics[width=0.48\textwidth]{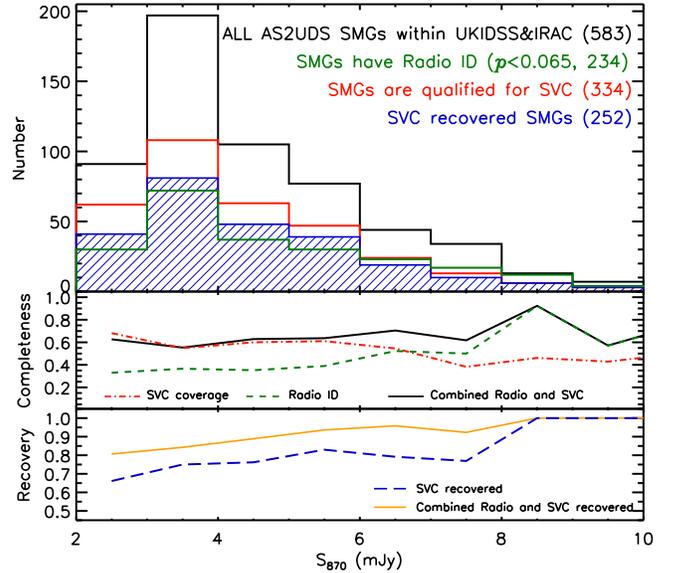}
\caption{The recovery rate and completeness of our combined radio and SVC  machine-learning methodology for identifying SMGs' counterparts as a function of flux density of SCUBA-2-detected submillimeter sources ($S_{\rm 870\mu m}$). 
We limit our identification of counterparts to single-dish-detected submillimeter sources in the combined region of UKIDSS and IRAC  since only $K$-band galaxies within this coverage are suitable for the machine-learning method. For the SMGs which have secure measurements of the five features used to train the SVC, our method successfully recovers 85 percent of SMGs, of these 77 percent can be recovered with just the machine-learning and this fraction increases to 81 percent for those submillimeter sources brighter than 4.5\,mJy at 870\,$\mu$m. For the full sample of AS2UDS SMGs within the combined UKIDSS and IRAC  coverage (not just those with the five features), 40 percent of SMGs have radio counterparts, and this fraction increases to 46 percent for brighter submillimeter sources ($S_{\rm 870\mu m} > $\,4.5\,mJy). Around 57 percent SMGs have the five features we use and so are qualified for our machine-learning method, and this fraction does not depend upon their submillimeter flux. Combining the radio identification and the machine-learning results, shows that 60 percent of ALMA-detected SMGs can be recovered and this fraction increases to 71 percent for the brighter submillimeter sources.}
\label{f:completeness.eps}
\end{figure}

It may be that the lower success rate for the ALESS sample is simply due to small number statistics: the number of ALMA maps in ALESS is only 12 percent (88/716) of that in AS2UDS. We can test this using AS2UDS, by selecting test samples of galaxies from 88 randomly selected ALMA maps from the AS2UDS survey and determining the variation in the recovery rate and precision between these test samples. We call this the ``12 percent test'' and we repeat this test 100 times to obtain the scatter. The median recovery rate of our combined radio and machine-learning method for a sample of galaxies in 88 ALMA maps is 86\,$\pm$\,5 percent with a lower-limit on the precision of 63\,$\pm$\,7 percent (Figure~\ref{f:ecdfs1.eps}). When we compare the results with the ``half--half'' test, we find that the smaller sample size causes larger uncertainties in the machine-learning classification. We also note that four of these 100 tests have the recovery rate as low as that of ALESS SMGs but with a range of  precisions. Therefore, it appears that the recovery rate and precision as low as those seen for the ALESS test sample are possible, just simply due to the small sample size. However, we note that there are also potential astrophysical reasons for the different success rates. In particular, the ALESS SMGs are typically fainter at 870\,\mm, median flux density of $S_{\rm 870\mu m}=$\,2.2\,mJy compared to AS2UDS SMGs, $S_{\rm 870\mu m}=$\,3.8\,mJy. As shown in Figure~\ref{f:completeness.eps}, the recovery rate of the combined radio and machine-learning method is higher for brighter SMGs. And we note that the beam of LABOCA/APEX, which is the basis of ALESS, is larger than that of SCUBA-2/JCMT used for AS2UDS, which will reduce the precision of identification of counterparts to single-dish-detected submillimeter sources.

Thus we argue that the decrease in the recovery rate when using the AS2UDS training set applied to the ALESS sample is probably partly caused by the relative faintness of the SMGs and larger beams of the single-dish survey in the ECDFS field. The difference between a $K$-selected training set in UDS and the IRAC-selected test sample in ECDFS may also affect the results of our method. 

We also undertook these same tests but now using the XGBoost machine-learning classifier. The two machine-learning modules have a very similar performance on the ``half-half'' and ``12 percent'' tests, while the XGBoost classifier gives a marginally higher recovery rate (64 percent) with a relatively lower precision (65 percent) for the ALESS sample.

%
%
\begin{table*}
\centering
\caption{Summary of radio-detected or machine-learning-classified \& ALMA-faint galaxies} \label{tab:table3}
\begin{tabular}{l|c|c|c}
\hline
\hline
Type of ALMA maps &  All ALMA maps &Maps with ALMA ID & ``Blank-ALMA'' maps \\
\hline
SVC-classified \& ALMA faint & 126/607\tablenotemark{a} & 75/512 & 51/95 \\%
Radio-detected \& ALMA faint &137/714 & 84/606 & 53/108 \\
\hline
\end{tabular}
\baselineskip=-3pt
\tablenotetext{a}{Values report number of galaxies and the number of eligible ALMA maps available.}
\end{table*}

\section{Results and Discussion} \label{s:discussion}

To determine the completeness of our method for recovering ALMA SMGs, we first summarise in Table~\ref{tab:table2} the  three evaluation metrics: recovery rate, precision and false positive rate, from the machine-learning classification, and the radio combined machine-learning method for SMGs in both the training set and test samples. We note that only $K$-band galaxies within the combined coverage of UKIDSS and IRAC  have sufficient photometric coverage to be suitable for the machine-learning method. Hence, the completeness was defined as the ratio between the number of recovered SMGs and the total number of ALMA SMGS within this overlapped region (\S~\ref{s:identification}).

We also report the number of radio-detected galaxies that are located within ALMA maps, but that do not have a $>$\,4.3\,$\sigma$ ALMA detection in Table~\ref{tab:table3}. We refer to these radio sources  as ``radio-detected \& ALMA faint'' galaxies in the following analysis since our stacking analysis shows that they are typically just below the submillimeter detection limit of our ALMA maps. For the same reason, the $K$-band galaxies which are classified as ``SMGs'' by the SVC machine-learning, but do not have a secure ALMA detection are termed ``machine-learning-classified \& ALMA faint'' galaxies. To verify that on average there is fainter submillimeter emission from galaxies within the ``blank-ALMA'' maps, as suggested by their detection in the far-infrared stacking analysis in \S\ref{s:blank-alma}, we separately study the properties of radio-detected/machine-learning-classified \& ALMA faint galaxies within the ``blank-ALMA'' maps and those maps which contain at least one ALMA-identified SMGs and list the number of each of them in Table~\ref{tab:table3}.

%
%

\subsection{Incompleteness of our multi-wavelength IDs}
We first investigate the completeness of our method for recovering ALMA SMGs in the UDS. We show the recovery rate of the combined radio and SVC machine-learning method as a function of flux density of SCUBA-2 detected submillimeter sources ($S_{\rm 870\mu m}$) in Figure~\ref{f:completeness.eps}. For the 583 AS2UDS SMGs within the overlapped region of UKIDSS and IRAC , 352/583 (60 percent) can be recovered by the combined radio and machine-learning method and for those submillimeter sources brighter than 4.5\,mJy at 870\,$\mu$m, this fraction increases to 71 percent. Of the 583 ALMA-identified SMGs, 334 have secure measurements of five selected properties and therefore qualify for the SVC machine-learning method. The SVC successfully selected 75 percent of these ALMA SMGs from a sample of all $K$-band-detected galaxies within the ALMA primary beams. By including the radio identifications, the recovery rate increase to 85 percent (285/334). 

Looking at the full SMG sample from AS2UDS within the combined UKIDSS and IRAC  coverage, the radio identification alone can recover 234/583 (40 percent) of all ALMA-detected SMGs, and for the brighter single-dish-detected submillimeter sources ($S_{\rm 850\mu m} \ge $\,4.5\,mJy), the recovery rate increases to 49 percent. For the AS2UDS SMGs that do not have secure measurement of five properties within the overlap-region of the UKIDSS and IRAC observations (and which we therefore cannot apply the SVC machine-learning method to), radio identification recovered an additional 67 SMGs. For the other 231/583 (40 percent) ALMA SMGs which are neither qualified for the SVC machine-learning method nor have radio counterparts, it is infeasible to identify their multi-wavelength counterparts. This fraction reduces to 29 percent for just those submillimeter sources brighter than 4.5\,mJy.

The purpose of this work is to construct a training set based on a large sample of ALMA-detected SMGs and deep ancillary data in the UDS field which can then be used to identify counterparts to single-dish-detected submillimeter sources from surveys of other fields which either have not yet been observed by ALMA or cannot be observed. Therefore, as a more representative test we also determine  the completeness and recovery rate of our method when applied to independent test samples: to separate halves of the AS2UDS survey sample and to the ALESS survey. For the ``half--half'' test on  AS2UDS, applying a training set constructed from half the AS2UDS maps to the galaxies in the other half of the maps,  the machine-learning recovers (75\,$\pm$\,3) percent of SMGs with a lower limit on the precision of 64\,$\pm$\,4 percent. When combined with the radio identifications, (86\,$\pm$\,3) percent of SMGs in the ``half--half'' AS2UDS test are recovered with a precision $>$\,61\,$\pm$\,3 percent. For the ALESS SMGs, 47/96 (49 percent) of ALESS SMGs are qualified for the machine-learning method and 29 of them were recovered. By including the radio identification, we can identify counterparts for 34/47 SMGs (72 percent). The radio identification recovers another 21 ALMA-detected SMGs by using the radio catalog from \cite{Biggs11} \citep{Hodge13, Simpson14}. In terms of the full-sample of ALESS SMGs, the combination of radio and machine-learning methods yields identifications for 55/96 (57 percent) of the counterparts to single-dish-detected submillimeter sources in the ECDFS.

One of the limitations that causes incompleteness in our method is the fact that SVC machine-learning method cannot deal with missing features unless they are artificially filled as we described in \S~\ref{s:machine-learn}. To test the affect of this limitation, we adopted the second machine-learning model, XGBoost, which has capacity of performing classifications with the missing values. The two machine-learning algorithms have very similar performances as we show in Table \ref{tab:table2}. The sample size for the machine-learning analysis is enlarged by including objects that lack $J$-band or 4.5\mm\ detection. This improve the completeness of analyses from 60 percent to 64 percent. However, for the other 46 percent AS2UDS SMGs, which are only detected in the submillimeter, or which have only counterparts in one or two wavebands, the opportunities to learn more about their properties even if they are correctly identified is limited due to the paucity of information available on them.

%
%
\begin{figure*}
\centering
\includegraphics[width=0.98\textwidth]{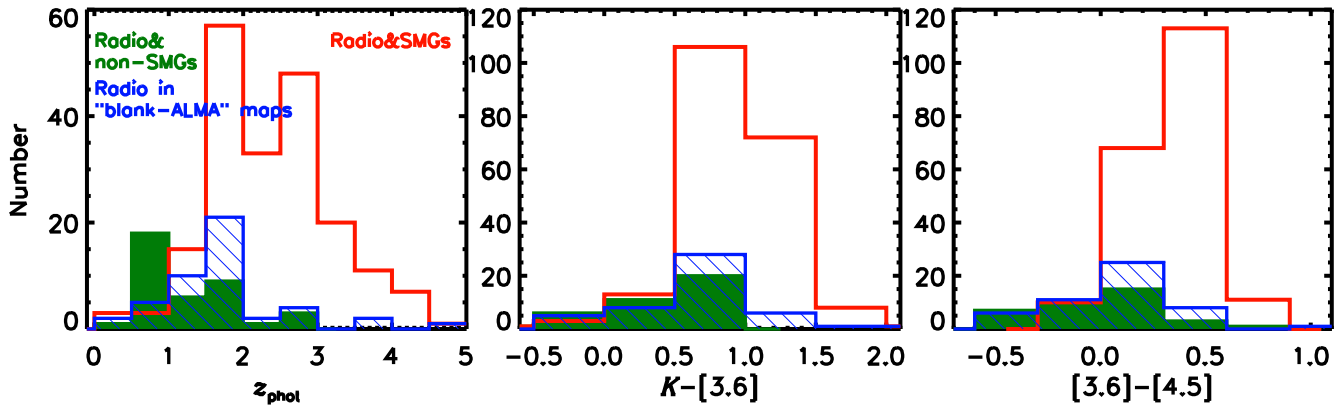}
\caption{The distributions of photometric redshift and near-infrared colors for those radio sources that have submillimeter counterparts (SMGs) versus those  that are not individually detected by ALMA (non-SMGs). Of the 137 non-SMG radio sources lying within our ALMA coverage, 53 are located in the ``blank-ALMA'' maps. We show the distribution of non-SMG radio sources within those maps which contain ALMA-identified SMG (green filled region) and those radio sources within the ``blank-ALMA'' maps (blue shaded region) respectively. The non-SMG radio sources lying in maps with an ALMA-detected SMG  tend to lie  at lower redshift and have bluer near-infrared colors. The distributions of radio sources within the ``blank-ALMA'' maps suggest that they may contain a mix of both higher-redshift, submillimeter sources  and lower-redshift, non-SMGs, which is consistent with our stacking results as we show later. }
\label{f:radio_smg_nsmg.eps}
\end{figure*}

%
%
\begin{figure}
\centering
\includegraphics[width=0.495\textwidth]{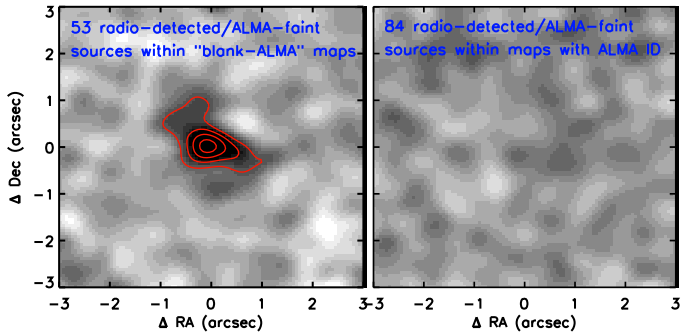}
\caption{The results of stacking the primary-beam-corrected ALMA maps at the position of those radio sources that are individually undetected at 870\,\mm\ with ALMA. The left panel shows the average stacking results of the 53 radio sources which are in ``blank-ALMA'' maps (those with no detected SMGs). The median peak flux density of these galaxies is $S_{\rm 870\mu m}=(0.51\pm0.05)$\,mJy. Contours indicate the significance of the 870\,\mm\  emission at 3, 6, 8, 10$\,\sigma$. The right panel is the similar stacked emission from the 84 non-SMG radio sources, but now lying in the maps which contain at least one ALMA-detected SMG. The stack results confirms that these radio sources are not submillimeter sources with a 3\,$\sigma$ limit of $S_{\rm 870\mu m}=$\,0.14\,mJy.  However, for the radio sources within ``blank-ALMA'' maps, at least, some of the radio sources are submillimeter sources, although these are just too faint (or too diffuse) to be individually detected in our ALMA observations.}
\label{f:stacking_maps_radio.eps}
\end{figure}

%
%
\begin{figure*}
\centering
\includegraphics[width=0.99\textwidth]{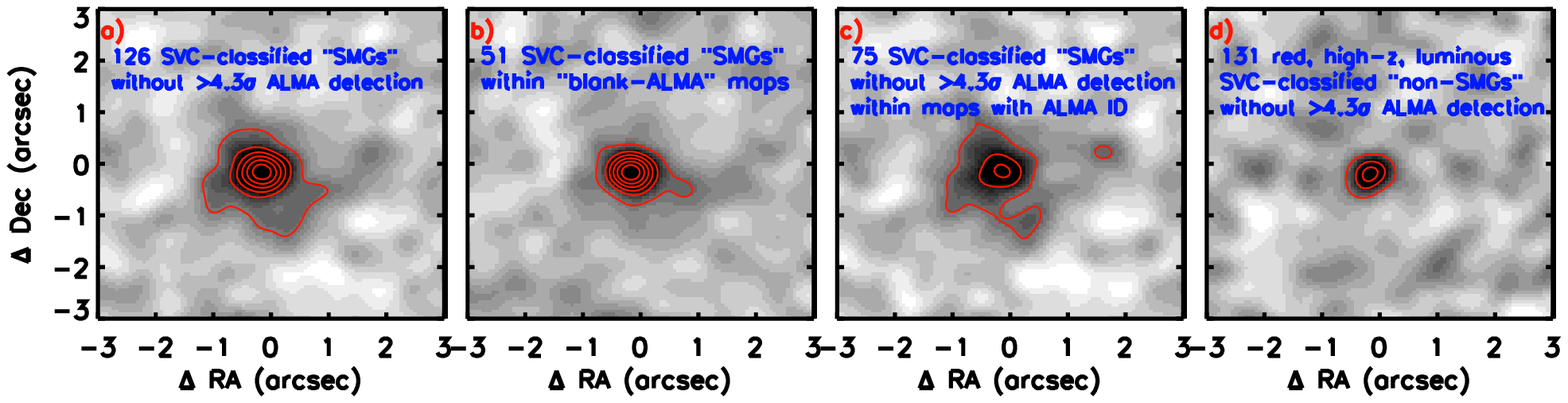}
\caption{The results of stacking the primary-beam-corrected ALMA maps at the position of $K$-band galaxies which are classified as ``SMG'' counterparts by the  machine-learning method, but which are not individually detected above $>$\,4.3\,$\sigma$ in our ALMA maps. {\bf a)} The average stacking results of all 126 such galaxies. We measure a median peak flux density of $S_{\rm 870\mu m}=(0.61\pm0.03)$\,mJy. Contours represent significance levels of 3, 6, 8, 10, 12, 14$\,\sigma$ at 870\,\mm. {\bf b)} The stacking results of the 51 machine-learning-classified \& ALMA faint galaxies which fall within ``blank-ALMA'' maps. The median peak flux for these galaxies is $S_{\rm 870\mu m}=(0.92\pm0.05)$\,mJy. {\bf c)} The averaged stacking results of the other 75 machine-learning-classified \& ALMA faint galaxies within those ALMA maps which contain a detected SMG. The median peak flux density for these sources is $S_{\rm 870\mu m}=(0.42\pm0.04)$\,mJy. Therefore, at least, on average the stacking results confirm that these machine-learning method-classified ``SMGs'' have detectable submillimeter emission at the $\sim$\,0.5--1\,mJy level. Equally interestingly, the median flux density of the machine-learning ``SMGs''  within ``blank-ALMA'' maps is twice that of similar ``SMGs'' in non-blank ALMA maps, which confirms that the SCUBA-2 detections in these regions are likely to be real (even if no individual galaxy was detectable with ALMA). {\bf d)} The stacking results of 131 redder ($(J-K)>2.0$, $(K-[3.6])>0.5$, $([3.6]-[4.5])>0.05$), high-redshift ($z_{phot}>$\,1.5), brighter ($M_{H}<-22.0$\,mag) $K$-band galaxies which are classified as ``non-SMGs'' by machine-learning. The stacking result shows that, on average, there is also fainter submillimeter emission from these galaxies. However, the peak flux density of this stacked map is $S_{\rm 870\mu m}=(0.30\pm0.03)$\,mJy which is half of the peak flux density of stacked maps at the position of 126 machine-learning classified ``SMGs'' (which also do not have a $>4.3\,\sigma$ALMA detection). This confirms that the machine-learning tends to pick out the brighter submillimeter galaxies. 
 }
\label{f:stacking_maps.eps}
\end{figure*}

\subsection{Properties of ALMA-detected and ALMA-faint radio sources} 
We investigate the population of radio-detected \& ALMA faint galaxies by comparing the multi-wavelength properties of ALMA-detected and ALMA-faint radio sources in UDS. The radio imaging covers 714/716 ALMA pointings. In total, 404 radio sources fall inside the 714 ALMA primary beams.  Among these, 259 match to ALMA SMGs within 1$\farcs$6, hence, are counterparts of ALMA SMGs. We define 137 radio sources as ``non-SMGs'' since they are located within the ALMA primary beams but $>$\,2$\farcs$6 away from ALMA SMGs or are within the ``blank-ALMA'' maps (eight radio sources lie between 1$\farcs$6--2$\farcs$6 from ALMA SMGs and are excluded from this analysis as their associations are ambiguous). We show the comparison of the multi-wavelength properties of these two samples of submillimeter-detected/undetected radio sources  in Figure~\ref{f:radio_smg_nsmg.eps}. We present the distribution of non-SMG radio sources lying in ALMA maps with a detected SMG and radio sources within the ``blank-ALMA'' maps separately, since the far-infrared stacking analysis shows that there may fainter submillimeter emissions from galaxies within the ``blank-ALMA'' maps (\S\ref{s:blank-alma}). The non-SMG radio sources within maps with an ALMA SMG tend to lie at lower redshift and are bluer in their near-infrared colors than the SMGs, i.e., they have the same properties as $K$-band-detected non-SMGs. This also confirms that our selected properties for the machine-learning can efficiently separate SMGs from field galaxies. Many of the radio sources within the ``blank-ALMA'' maps have properties like SMGs, while some show the properties of non-SMGs.

To further investigate the radio-detected, but ALMA-undetected, galaxies in our field, we stack the primary-beam-corrected ALMA maps at the position of these radio sources. There are 404 radio sources located within the $\sim$\,50\,arcmin$^2$ covered by our ALMA survey. Among these, 137 are defined as non-SMGs since they do not correspond to a $>$\,4.3\,$\sigma$ ALMA counterpart. We separately stack the 53 radio sources that lie in ``blank-ALMA'' maps and 84 non-SMG radio sources  in maps with at least one ALMA-detected SMG. We show the stacked results of 53/137 radio sources within the ``blank-ALMA'' maps in the left panel of Figure~\ref{f:stacking_maps_radio.eps}. The median  flux density of the stacked ALMA images is $S_{\rm 870\mu m}=(0.51\pm0.05)$\,mJy which is consistent with the detection of significant far-infrared emission in the SPIRE stacks of these maps (\S\ref{s:blank-alma}), indicating that most of them correspond to real SCUBA-2 sources. We also stack the 84 radio sources which are individually undetected by ALMA, but lie in a map with an ALMA-detected SMGs. This confirms that these galaxies do not have detectable submillimeter emission, i.e., they are non-SMGs. Based on the fraction of SMGs and non-SMGs in the maps with an ALMA-identified SMG, we estimate that at least $\sim$\,70 percent (Figure~\ref{f:radio.eps}) of radio sources in the ``blank-ALMA'' maps have real submillimeter emission, although, they are too faint to be detected individually in our ALMA observations.

\subsection{Stacking machine-learning-classified \& ALMA faint $K$-band galaxies}
As shown in Figure~\ref{f:self-test.eps}, the machine-learning-classified \& ALMA faint $K$-band galaxies have similar properties to ALMA-detected SMGs: they lie at high redshift, they are bright in the rest-frame $H$-band and red in optical/near-infrared colors. 

To determine whether these galaxies are submillimeter-emitters that lie slightly below the detection limit of our 870\,\mm\ ALMA maps, we perform a stacking analysis at their positions in the ALMA maps. In Figure~\ref{f:self-test.eps}, we show that the machine-learning method classifies 378 ``SMGs'' from the 2033 $K$-band galaxies within the ALMA primary beams. Among these, 252 match to ALMA-identified SMGs. We show the stacked results of the other 126 SVC-classified, but ALMA-undetected, $K$-band galaxies in Figure~\ref{f:stacking_maps.eps}. The median flux density of these 126 galaxies is $S_{\rm 870\mu m}=(0.61\pm0.03)$\,mJy, which indicates that on average these sources have submillimeter emission, but are too faint (or too diffuse) to be detected by our ALMA observations. 

Among these 126 $K$-band galaxies, 51 of them lie in the ``blank-ALMA'' maps (those without an individually detected SMG). The stacked median flux density of these 51 galaxies is $S_{\rm 870\mu m}=(0.92\pm0.05)$\,mJy. The other 75 galaxies,  those in maps with an ALMA-identified SMG, have stacked median flux density of $S_{\rm 870\mu m}=(0.42\pm0.04)$\,mJy (Figure~\ref{f:stacking_maps.eps}). Therefore, the machine-learning classified ``SMGs'' within the ``blank-ALMA'' maps have submillimeter emission that is twice as bright as similar galaxies in those maps that already contain an individually-detected SMG. This confirms our suggestion that  some galaxies within the ``blank-ALMA'' maps have submillimeter emission just below the detection threshold of our ALMA observations. Therefore, because of the ambiguous classification of the sources within the ``blank-ALMA'' maps, we chose not to include them when originally constructing the training sets for the machine-learning (\ref{s:machine-learn}), as they would have blurred the distinction between the properties of the SMGs and non-SMGs. We also investigate the effect on the machine-learning training by including $K$-band galaxies within the ``blank-ALMA'' maps into the non-SMG sample of the training set. In this case, the recovery rate of SMGs based on the ``self-test'' decreases by about 10 percent. The reason for this is that in this case we are labelling some  galaxies which have the same properties as counterparts of ALMA SMGs as ``non-SMGs''. 

We also compare the machine-learning results to that of simple cuts on the near-infrared colors, photometric redshifts, and absolute $H$-band magnitudes to select the redder (($(J-K)>2.0$, $(K-[3.6])>0.5$, $([3.6]-[4.5])>0.05$)), higher-redshift ($z_{phot} >$\,1.5), and brighter ($M_{H} < -22.0$\,mag) galaxies as probable counterparts of SMGs. There are 483 $K$-band galaxies in the AS2UDS test sample that meet the above criteria. Among these, 251 are ALMA-detected SMGs. Therefore, the recovery rate of this simple method is similar to that of the machine-learning method (which recovers 252 ALMA-detected SMGs). However, the precision of this simple method is just 52 percent (251/483), 15 percent lower than that of the machine-learning method.   Thus while simple cuts on a small number of observables can be used to identify probable SMG counterparts, the contamination from non-submillimeter-bright galaxies is significantly worse than that achieved by the machine-learning method. In fact from the 483 redder, higher-redshift, and brighter galaxies selected by these simple parametric cuts, 131 are classified as ``non-SMGs'' by the machine-learning method.   Stacking the primary-beam-corrected ALMA maps at the position of these 131 $K$-band galaxies, we find that the peak flux density of the stacked map is $S_{\rm 870\mu m}=(0.30\pm0.03)$\,mJy. This is half of the peak flux density of the stacked maps at the positions of the 126 machine-learning classified ``SMGs'' (which also do not have an $>4.3\,\sigma$ ALMA detection, Figure~\ref{f:stacking_maps.eps}). This confirms that the machine-learning approach is more effective than simple cuts in terms of identifying the counterparts of brighter submillimeter sources.

\section{Conclusions}\label{s:conclusion}

From ALMA follow-up observations of the 716 SCUBA-2 detected submillimeter sources in the S2CLS UKIDSS-UDS field \citep{Stach18a,Stach18b}, we exploit a sample of 695 submillimeter galaxies ($>$\,4.3\,$\sigma$) within 608 ALMA maps. We label the other 108 ALMA maps which do not contain a $>$\,4.3\,$\sigma$ ALMA SMGs as ``blank-ALMA'' maps. Utilising our high-resolution ALMA data, we first identify radio, optical, and near-infrared counterparts to the ALMA SMGs. We define as a ``non-SMG'' any radio/$K$-band galaxies that are located within the primary beams of our ALMA maps but are $>$\,2$\farcs$6 (radio non-SMGs) or $>$1$\farcs$6 ($K$-band non-SMGs) away from an ALMA-detected SMGs. 
Based on the samples of ALMA SMGs and non-SMGs, we develop a  combined radio and machine-learning method using Support Vector Classification to identify multi-wavelength counterparts to the single-dish-detected submillimeter sources. The main conclusions from our work are as follows:

1. We identify radio counterparts to the  ALMA SMGs in the UDS. In total, there are 404 radio sources within the primary beam coverage of our ALMA maps. Out of 695 ALMA SMGs, 268 match to 259 radio sources within 1$\farcs$6. We adopt a $p$-value cut to identify radio counterparts to single-dish submillimeter sources. We consider 363 of the 404 radio sources with $p<$\,0.065 as counterparts to SMGs. Among them, 254 are matched to 263 ALMA SMGs within 1$\farcs$6. The radio identification step can recover 37 percent of SMGs from the single-dish survey in UDS with a precision $>$\,70 percent.

2. We identify optical/near-infrared counterparts by matching the ALMA SMGs to a deep $K$-band-detected photometric catalog. Within the overlap region of UKIDSS and IRAC  coverage, 483 $K$-band galaxies match to ALMA SMGs within 0$\farcs$6. Therefore, $\sim$\,83 percent (483/583) of the ALMA SMGs in this region have $K$-band counterparts. We find that the photometric redshift, absolute rest-frame $H$-band magnitude, and near-infrared colors ($(J-K)$, $(K-[3.6])$ and ([3.6]-[4.5])) of these SMGs appear to provide the most diagnostic power to  differentiate SMGs from non-SMGs. We construct a training set that includes ALMA SMGs and non-SMG $K$-band galaxies with  secure measurements of these five selected properties. We do not include those $K$-band galaxies within the ``blank-ALMA'' maps in the non-SMG sample used in the training set since our stacking results indicate that these sources are faint submillimeter emitters. 

3. We train the SVC machine-learning classifier using a training set of SMGs and non-SMGs and then classify the sources in a test sample. We perform a ``self-test'' of our machine-learning method by classifying all of the 2033 $K$-band galaxies that have secure measurements of the five selected properties and which within the ALMA primary beams in the UDS. Among these, 334 are AS2UDS ALMA-detected SMGs. The machine-learning classifies 378 $K$-band galaxies as the counterparts of SMGs with a recovery rate of 75 percent and a precision of 67 percent. Our stacking results show that there is faint submillimeter emission which is just below our ALMA detection threshold from the galaxies which are classified as ``SMGs'' (but are not ALMA detected) by the SVC machine-learning method. Therefore, both the recovery rate and precision of the machine-learning method should be considered as the lower limits. Combined with the radio identification, our method can recover $>$\,85 percent SMGs which have secure measurements of five selected features with a precision of $>$\,62  percent. 

4. To test our method we use a training set constructed from the galaxies in a randomly selected half of our AS2UDS ALMA maps to an independent test sample from the other half of the ALMA maps. We estimate a recovery rate of 86\,$\pm$\,3 percent and a precision of $>$\,61\,$\pm$\,3 percent from this ``half--half'' test, confirming the robustness of our method of identifying counterparts for single-dish-detected submillimeter sources when using a training set from the same field. We also apply our method from a $K$-detected training set in the UDS field to the IRAC-detected galaxies in the ECDFS field to predict counterparts to LABOCA-detected submillimeter sources. We use the ALMA-detected SMGs in this field from the ALESS survey to check the recovery rate and precision of our method. The combined radio and machine-learning method recovers 72 percent of ALMA SMGs with a lower limit on the precision of 65 percent. We show that the decrease of recovery rate is likely to be  partly the relative faintness of the ALESS SMGs and larger beam of APEX/LABOCA, compared to those in the AS2UDS training set. The difference between $K$-band-detected training set in the UDS field and IRAC-detected test sample in the ECDFS field may also affect the precision of our method. We also show that the smaller sample size of ALESS causes increased uncertainties in the classifications.

5. The main limitation of our method is that we miss those SMGs that do not qualify for the machine-learning and do not have a radio counterparts. We estimate the fraction of missed sources by checking the recovery rate of ALMA SMGs. In the overlapped region of ALMA, UKIDSS, IRAC and VLA, 60 percent of ALMA SMGs in the UDS field are recovered by our combined radio and machine-learning method. This fraction increases to 71 percent for SMGs brighter than $S_{\rm 850\mu m}\geq $\,4.5\,mJy. The completeness of recovered SMGs increases to 64 percent if we adopt a second machine-learning module, XGBoost. This machine-learning algorithm has a very similar performance as SVM in classifying SMGs from $K$-band detected field galaxies but can deal with  missing features and so can employ a slightly large test sample by including objects with only limits on their $J$-band or 4.5\,\mm\ fluxes.

6. By stacking the emission in ALMA maps at the position of the machine-learning classified, but individually ALMA-undetected, $K$-band galaxies we show that on average there is faint submillimeter emission from these galaxies. Moreover, a stack of the far-infrared {\it Herschel} SPIRE maps at the position of the ``blank-ALMA'' maps and a stack of ``blank-ALMA'' maps at the position of machine-learning classified ``SMGs'' within these maps demonstrate that the majority of SCUBA-2 sources are real, although the submillimeter galaxies responsible for these sources are either too faint and/or diffuse to be detected by ALMA.

In summary,  the combined radio and machine-learning technique developed in this work can be used to construct large samples of likely SMG counterparts from wide-field single-dish submillimeter surveys which currently lack interferometric submillimeter follow-up, such as the remaining fields in S2CLS \citep{Geach17} or S2COSMOS \citep{Simpson18}. These statistically large samples will enable us to investigate science questions related to the formation of SMGs, their evolutionary connections with other populations, such as high-redshift QSOs, and compact, red galaxies at $z\sim $\,1--3 and ultimately massive galaxies at $z\sim$\,0.

We publish the training set of SMG and non-SMG sources from the AS2UDS survey as a machine-readable catalog with this paper to allow others to apply the machine-learning method we adopted in this work to other fields.

\acknowledgments
We are grateful to the referee and statistics editor for a detailed report and valuable comments, which have improved the quality of this work. FXA acknowledges support from the China Scholarship Council for studying two years at Durham University. All Durham co-authors acknowledge STFC support through grant ST/P000541/1. IRS, BG, and EAC acknowledge the ERC Advanced Grant DUSTYGAL (321334). IRS also acknowledges a Royal Society/Wolfson Merit Award. FXA also acknowledges support from the National Key Research and Development Program of China (No.\ 2017YFA0402703) and NSFC grant  (11773076). JLW acknowledges the support of an Ernest Rutherford Fellowship. JEG acknowledges Royal Society. F.X.A. acknowledges Ryley Hill for helpful discussion about the machine-learning algorithms. MJM acknowledges the support of the National Science Centre, Poland through the POLONEZ grant 2015/19/P/ST9/04010; this project has received funding from the European Union's Horizon 2020 research and innovation programme under the Marie Sk{\l}odowska-Curie grant agreement No. 665778. We thank the staff at UKIRT for their efforts in ensuring the success of the UDS project. The James Clerk Maxwell telescope has historically been operated by the Joint Astronomy Centre on behalf of the Science and Technology Facilities Council of the United Kingdom, the National Research Council of Canada, and the Netherlands Organisation for Scientific Research. Additional funds for the construction of SCUBA-2 were provided by the Canada Foundation for Innovation. This paper makes use of the following ALMA data: ADS/JAO.ALMA\#2012.1.00090.S, 2015.1.01528.S, and 2016.1.00434.S. ALMA is a partnership of ESO (representing its member states), NSF (USA), and NINS (Japan), together with NRC (Canada), NSC and ASIAA (Taiwan), and KASI (Republic of Korea),in cooperation with the Republic of Chile. The Joint ALMA Observatory is operated by ESO, AUI/NRAO, and NAOJ.

\end{document}